\documentclass{jkas}

\def\year{2023} % publication year
\def\volume{56} % journal volume
\def\issue{1} % journal issue
\def\beginpage{1} % first page of article
\def\received{---} % date paper was received by JKAS
\def\accepted{---} % date of acceptance
\def\published{---} % date of publication
\date{Received \received; Accepted \accepted; Published \published}

\usepackage{orcidlink}
\usepackage{gensymb}
\usepackage{color, soul}
\usepackage{longtable}
\usepackage{ulem}
\title{%
Interferometric Monitoring of Gamma-ray Bright AGNs: 4C~+28.07 and its Synchrotron Self-absorption Spectrum
}

\author[1]{Myoung-Seok Nam}{0009-0001-4748-0211}
\author[1, 2, $\star$]{Sang-Sung Lee}{0000-0002-6269-594X}
\author[1,2]{Whee Yeon Cheong}{0009-0002-1871-5824}
\affil[1]{Korea Astronomy and Space Science Institute, Daedeok-daero 776, Yuseong-gu, Daejeon 34055, Republic of Korea}
\affil[2]{University of Science and Technology, Gajeong-ro 217, Yuseong-gu, Daejeon 34113, Republic of Korea}
\def\corrauthor{%
Sang-Sung Lee, \email{sslee@kasi.re.kr}
}

\def\runningauthor{%
Nam et al.
}

\def\runningtitle{%
iMOGABA: 4C +28.07
}

\def\keywords{%
galaxies: active --- galaxies: jets --- galaxies: nuclei --- galaxies: individual: 4C~+28.07
}

\def\abstracttext{%
We present the analysis results of the simultaneous multifrequency observations of the blazar 4C~+28.07. The observations were conducted by the Interferometric Monitoring of Gamma-ray Bright Active Galactic Nuclei~(iMOGABA) program, which is a key science program of the Korean Very Long Baseline Interferometry~(VLBI) Network~(KVN). Observations of the iMOGABA program for 4C~+28.07 were conducted from 16 January 2013~(MJD~56308) to 13 March 2020~(MJD~58921). We also used ${\gamma}$-ray data from the Fermi Large Array Telescope~(Fermi-LAT) Light Curve Repository, covering the energy range from 100~MeV to 100~GeV. We divided the iMOGABA data and the Fermi-LAT data into five periods from 0 to 4, according to the prosody of the 22~GHz data and the presence or absence of the data. In order to investigate the characteristics of each period, the light curves were plotted and compared. However, a peak that formed a hill was observed earlier than the period of a strong ${\gamma}$-ray flare at 43-86~GHz in period~3~(MJD~57400-58100). Therefore, we assumed that the minimum total CLEANed flux density for each frequency was quiescent flux~(${S_{\rm{q}}}$) in which the core of 4C~+28.07 emitted the minimum, with the variable flux~(${S_{\rm{var}}}$) obtained by subtracting ${S_{\rm{q}}}$ from the values of the total CLEANed flux density. We then compared the variability of the spectral indices~(${\alpha}$) between adjacent frequencies through a spectral analysis. Most notably, ${\alpha_{\rm{22-43}}}$ showed optically thick spectra in the absence of a strong ${\gamma}$-ray flare, and when the flare appeared, ${\alpha_{\rm{22-43}}}$ became optically thinner. In order to find out the characteristics of the magnetic field in the variable region, the magnetic field strength in the synchrotron self-absorption~(${B_{\rm{SSA}}}$) and the equipartition magnetic field strength~(${B_{\rm{eq}}}$) were obtained. We found that $B_{\rm SSA}$ is largely consistent with $B_{\rm eq}$ within the uncertainty, implying that the SSA region in the source is not significantly deviated from the equipartition condition in the ${\gamma}$-ray quiescent periods.
}

\begin{document}
%\jkashead %% set title, authors, abstract, etc.

%%%%%%%%%%%%%%%%%%%%%%%%%%%%%%%%%%%%%%%%%%%%%%%%%%%%%%%%%%%%%%%%%%%%%
%%% BEGIN MAIN TEXT HERE %%%%%%%%%%%%%%%%%%%%%%%%%%%%%%%%%%%%%%%%%%%%
%%%%%%%%%%%%%%%%%%%%%%%%%%%%%%%%%%%%%%%%%%%%%%%%%%%%%%%%%%%%%%%%%%%%%
\onecolumn
{\LARGE \textbf{Interferometric Monitoring of Gamma-ray Bright AGNs: 4C~+28.07 and its Synchrotron Self-absorption Spectrum}}
\\\\
{\large \textbf{Myoung-Seok Nam\orcidlink{0009-0001-4748-0211}$^1$, Sang-Sung Lee\orcidlink{0000-0002-6269-594X}$^{1, 2, \star}$, and Whee Yeon Cheong\orcidlink{0009-0002-1871-5824}$^{1, 2}$}}
\\
\textcolor{gray}{$^1$ Korea Astronomy and Space Science Institute, Daejeon 34055, Republic of Korea}
\\
\textcolor{gray}{$^2$ University of Science and Technology, Daejeon 34113, Republic of Korea}
\\
$^\star$ Corresponding Author: Sang-Sung Lee, sslee@kasi.re.kr
\\\\\\
\textbf{Abstract}
\\
We present the analysis results of the simultaneous multifrequency observations of the blazar 4C~+28.07. The observations were conducted by the Interferometric Monitoring of Gamma-ray Bright Active Galactic Nuclei~(iMOGABA) program, which is a key science program of the Korean Very Long Baseline Interferometry~(VLBI) Network~(KVN). Observations of the iMOGABA program for 4C~+28.07 were conducted from 16 January 2013~(MJD~56308) to 13 March 2020~(MJD~58921). We also used ${\gamma}$-ray data from the Fermi Large Array Telescope~(Fermi-LAT) Light Curve Repository, covering the energy range from 100~MeV to 100~GeV. We divided the iMOGABA data and the Fermi-LAT data into five periods from 0 to 4, according to the prosody of the 22~GHz data and the presence or absence of the data. In order to investigate the characteristics of each period, the light curves were plotted and compared. However, a peak that formed a hill was observed earlier than the period of a strong ${\gamma}$-ray flare at 43-86~GHz in period~3~(MJD~57400-58100). Therefore, we assumed that the minimum total CLEANed flux density for each frequency was quiescent flux~(${S_{\rm{q}}}$) in which the core of 4C~+28.07 emitted the minimum, with the variable flux~(${S_{\rm{var}}}$) obtained by subtracting ${S_{\rm{q}}}$ from the values of the total CLEANed flux density. We then compared the variability of the spectral indices~(${\alpha}$) between adjacent frequencies through a spectral analysis. Most notably, ${\alpha_{\rm{22-43}}}$ showed optically thick spectra in the absence of a strong ${\gamma}$-ray flare, and when the flare appeared, ${\alpha_{\rm{22-43}}}$ became optically thinner. In order to find out the characteristics of the magnetic field in the variable region, the magnetic field strength in the synchrotron self-absorption~(${B_{\rm{SSA}}}$) and the equipartition magnetic field strength~(${B_{\rm{eq}}}$) were obtained. We found that $B_{\rm SSA}$ is largely consistent with $B_{\rm eq}$ within the uncertainty, implying that the SSA region in the source is not significantly deviated from the equipartition condition in the ${\gamma}$-ray quiescent periods.
\twocolumn

\section{Introduction}\label{sec:1}

Active galactic nuclei~(AGNs) are the brightest and most dynamic objects in the universe~\citep{eht2019first}. 
AGNs are classified as seyferts, radio galaxies, quasars, and blazars~\citep{carroll2017introduction}, and mainly consisting of a supermassive black hole~(SMBH), an accretion disk, and relativistic jets. Generally, the mass of the supermassive black holes of AGNs is known to be ${10^{6}}$-${10^{10}~M_{\odot}}$~\citep{padovani2017active}. The relativistic jets are plasma-launched from the vicinity of the inner regions in AGNs, moving at relativistic speeds and showing apparent motion at speeds ranging from ${\sim 0.03~\rm{c}}$ to 40~c~\citep{kellermann2004sub, blandford2019relativistic}, and radiating a wide range of electromagnetic waves from radio to ${\gamma}$-ray. Synchrotron radiation is emitted from the jets due to relativistic charged particles~(e.g., electrons) accelerated by magnetic fields, yielding a wide frequency range~(from radio to X-ray) of continuum radiation. The relativistic particles in the jets may interact with photons around and up-scatter the photons with higher energy levels of, for instance, ${\gamma}$-ray~(inverse Compton scattering or IC). The seed photons for the scattering may come from the synchrotron radiation field~(Synchrotron Self-Compton or SSC) or may have an external origin~(external inverse Compton or EIC). It is also expected that low-energy synchrotron radiation is absorbed by dense particles~(most likely at low energy levels), also known as synchrotron self-absorption~(or SSA).

Blazars are a subclass of AGNs~\citep{angel1980optical} with two main groups: BL Lacertae objects~(BL Lacs) and flat-spectrum radio quasars~(FSRQs) according to~\citet{urry1995unified}. Blazars have certain characteristic properties, such as rapid variability, high degrees~(up to 20\%) of linear polarization at visible wavelengths~\citep{carroll2017introduction}, and small viewing angles ~(of a few degrees) between the relativistic jet and the line of sight~\citep{jorstad2017kinematics}. These properties make blazars compact with angular sizes in the sub-arcsecond~(e.g., milli-arcsecond or mas) range, giving them highly variable degrees of brightness with variable time scales of hours to years.

4C~+28.07 is a FSRQ~\citep{fey2004second}~(${z~=~1.206}$) with a SMBH mass of ${M_{\rm{BH}}~=~(1.65_{-0.82}^{+1.66}) \times 10^{9} \rm{M}_{\odot}}$~\citep{shaw2012spectroscopy, zargaryan2022multiwavelength}. Multiwavelength observational studies of 4C~+28.07 have mainly revealed the physical properties of $\gamma$-ray flares from the source. \citet{das2021multi} conducted a long-period cross-correlation study of 4C~+28.07, analyzing multi-wavelength ($\gamma$-ray to radio) data obtained from the Fermi-Large Area Telescope~\citep[Fermi-LAT;][]{atwood2009large}, Swift-XRT/UVOT~\citep{gehrels2004swift, burrows2005swift, roming2005swift}, Catalina surveys~\citep{drake2009first}, the Steward Optical Observatory~\citep{smith2009coordinated}, and from the Owens Valley Radio Observatory~\citep[OVRO;][]{richards2011blazars}. They found three distinctive flares~(MJD~54682-59000) in the Fermi-LAT data and reported that two of them in MJD~55746-56000 and MJD~58355-58574 are significantly correlated with the radio light curve with large delays of 70-135~days, concluding that the emitting regions of radio and ${\gamma}$-ray are separated each other. They found that a leptonic scenario suitably explains the two different emission regions in the source using a two-zone model for the broadband spectral energy distribution~\citep[SED;][]{abdo2010spectral} modeling. 

Another multiwavelength study of the source was performed by \citet{zargaryan2022multiwavelength}, using $\gamma$-ray, X-ray, ultraviolet, and optical data of 4C~+28.07 obtained from Fermi-LAT and Swift-XRT/UVOT. They found five flares in the ${\gamma}$-ray band, and using the Bayesian Block approach~\citep{scargle2013studies} analyzed the flares, in particular the strongest fifth flare on 12 October 2018~(MJD 58403), revealing that an EIC model is highly applicable to the fifth flare and that the SSC flare is applicable to the remaining period. Interestingly, they determined that the size of the ${\gamma}$-ray-emitting region is close to the Schwarzschild radius, mentioning 4C~+28.07 as an excellent laboratory for AGN physics.

Although the source has been well investigated for the correlations between the $\gamma$-ray flares and the multiwavelength variability, the spectral variations in radio-emitting regions have not been investigated for the $\gamma$-ray active periods. In this paper, multifrequency radio data obtained simultaneously at 22, 43, 86, and 129~GHz using the Korean very long baseline interferometry~(VLBI) Network~(KVN) are used to investigate the spectral variations of 4C~+28.07 during a $\gamma$-ray active period of the source and to study possible correlations between the radio flux density~(as well as the spectral properties) and the $\gamma$-ray flux density. In Section~\ref{sec:2}, we describe the multifrequency observations and the data used in this paper. In Section~\ref{sec:3}, the data reduction and imaging processes are introduced in detail, and in Section~\ref{sec:4}, the results are summarized. In Section~\ref{sec:5}, we discuss the analysis results. Finally, we summarize the preceding sections in Section~\ref{sec:6}.

\section{Observations and Data}\label{sec:2}
\subsection{KVN}
\subsubsection{Observations at 22, 43, 86, and 129~GHz}
The KVN was built by the Korea Astronomy and Space Science Institute~(KASI) in 2001 and became the first VLBI network in East Asia. The KVN consists of three 21-m diameter radio telescopes located in Seoul~(KVN Yonsei; KY), Ulsan~(KVN Ulsan; KU), and Jeju~(KVN Tamna; KT), yielding a baseline range of 300-480~km, which will be extended to 130-500~km with a fourth station, KVN PyeongChang~\footnote{\url{https://radio.kasi.re.kr/kvn/}}. The KVN enables us to run simultaneous multifrequency VLBI observations at four frequencies~(22, 43, 86, and 129~GHz), a capability that has motivated several KVN key science programs, including the interferometric Monitoring of Gamma-ray Bright AGNs~(iMOGABA).

The principal purpose of the iMOGABA program is to study the origins of the ${\gamma}$-ray flares of AGNs~\citep{lee2016interferometric}. The iMOGABA program has monitored approximately 30 sources since December of 2012~\citep{lee2017interferometric}, and the KVN observations are done in left circular polarization and with a total bandwidth of 256~MHz~(64~MHz for each of the four frequency bands). In order to achieve accurate amplitude calibrations, the atmospheric opacity has been measured every hour during the observations by measuring system temperatures at eight elevations~(i.e., Sky Dipping), and the antenna pointing direction has been corrected upon every scan, when available, by conducting cross-scan observations of each target source. 4C~+28.07 has been monitored as part of the iMOGABA program, yielding multifrequency~(22-129~GHz) simultaneous VLBI observations at an angular resolution range of 1-6 mas during the period of 16 January 2013~(MJD~56308) to 13 March 2020~(MJD~58921). The data observed from the iMOGABA program were recorded by the MARK~5B system~\citep{whitney2002mark, lee2015newhardware}. Afterwards, the data were correlated by the DiFX software correlator in Daejeon, Korea~\citep{deller2011difx, lee2016interferometric}.

\subsubsection{Data Reduction and Imaging}\label{sec:3} 

In the iMOGABA program, correlated data were reduced by the KVN pipeline~\citep{hodgson2016automatic} using the Astronomical Image Processing System~\citep[AIPS;][]{wells1985nrao}. The pipeline goes through the following steps: loading of FITS files~(FITLD), reading of amplitude calibration information~(ANTAB), application of TY and GC tables~(APCAL), amplitude correction via auto-correlation~(ACCOR), amplitude bandpass correction~(BPASS), separation of the dataset according to the frequency~(UVCOP), fringe fitting~(FRING), running of a ParselTongue script to resolve delays~\citep[IF-Aligner;][]{marti2012calibration}, performing a frequency phase transfer~\citep[FPT;][]{algaba2015interferometric}, separation of files for each source~(SPLIT), and saving of the separated files in the FITS format~(FITTP). The procedure of the KVN pipeline is similar to that of the pipeline of the European VLBI Network. Nevertheless, the pipeline of the KVN has two differences: the data go through an \texttt{IF-Aligner} procedure in the form of an automated manual phase calibration, with a subsequent \texttt{FPT} procedure using the simultaneous observation capabilities of the KVN~\citep[see][for more details]{hodgson2016automatic}. Consequently, the data obtained through the pipeline undergoes an imaging procedure.

\begin{figure*}[hbt!]
\centering
\includegraphics[width=\textwidth]{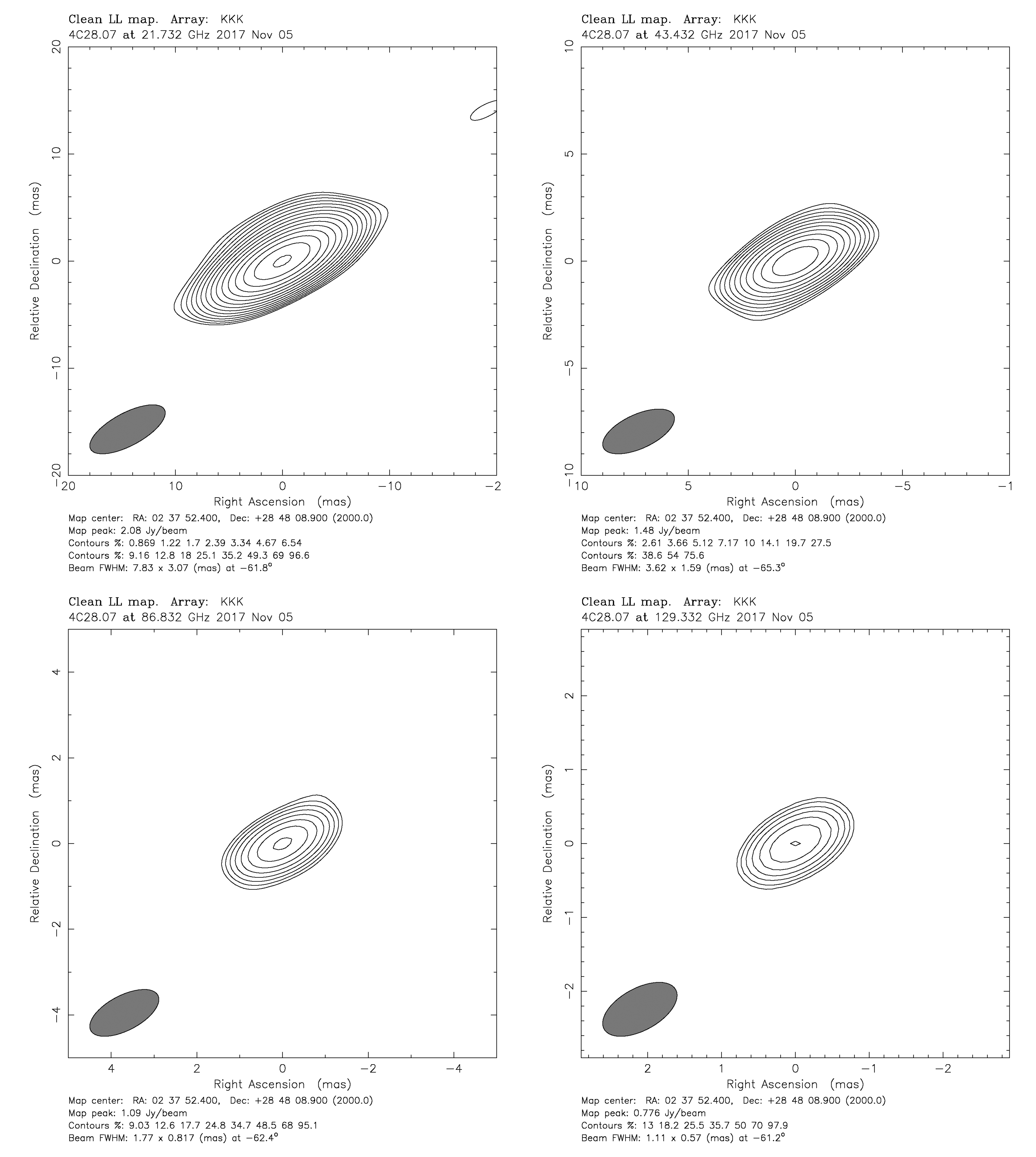}
\caption{The CLEANed images of 4C +28.07 were obtained from the iMOGABA observation on 2017 November 5. (Upper) 22~GHz on the left, 43~GHz on the right. (Lower) 86~GHz on the left, 129~GHz on the right. A CLEAN beam~(a grey ellipse) is shown in the lower left corner of each image. Peak flux density, contour levels~(in percent of the peak flux density), and the CLEAN beam information are provided at the bottom of each image. The lowest contour level is determined as $3\times \sigma_{\rm rms}$ of each map. The contours have a logarithmic spacing and are drawn at ${-1}$, 1, 1.4, ..., and $1.4^n$ of the lowest contour level.\label{fig:img_residual}}
\end{figure*}

We produced CLEANed images of 4C~+28.07 using the difference mapping~(DIFMAP) software developed at Caltech to display and edit the data and for self-calibration, imaging, and deconvolution~\citep{1994BAAS...26..987S}. The pipelined \textit{uv}-data are averaged over time for 30~s, 30~s, 20~s, and 10~s at 22, 43, 86, and 129~GHz, and the averaged data are investigated in \texttt{vplot} to flag amplitude values~(outliers) that significantly deviate from the flow of the amplitude. Multifrequency CLEANed images of 4C~+28.07 have been produced using a script~(see Appendix~\ref{appendix_imaging}), and the imaging parameters are summarized in Table~\ref{tab:cln_param}. An example of CLEANed images from 5 November 2017~(MJD~58062) is shown in Figure~\ref{fig:img_residual}, where the map peak fluxes are 2.08, 1.48, 1.09, and 0.776~Jy/beam, and the contour ranges for percent of the peak flux density are 0.869-96.6, 2.61-75.6, 9.03-95.1, and 13-97.9 at 22, 43, 86, and 129~GHz, respectively.

%%% FIGURE %%%%%%%%%%%%%%%%%%%%%%%%%%%%%%%%%%%%%%%%%%%%%%%%%%%%%%%%%%
%%%%%%%%%%%%%%%%%%%%%%%%%%%%%%%%%%%%%%%%%%%%%%%%%%%%%%%%%%%%%%%%%%%%%

%%% TABLE %%%%%%%%%%%%%%%%%%%%%%%%%%%%%%%%%%%%%%%%%%%%%%%%%%%%%%%%%%%
\onecolumn
\begin{table}[p]
\caption{\label{tab:cln_param} Image parameters of the CLEANed results using the script.}
\centering
\resizebox{\textwidth}{!}{%
{\scriptsize
\begin{tabular}{lcccccccccc}
\toprule
Date & MJD & Band & ${B_{\rm maj}}$ & ${B_{\rm min}}$ & ${B_{\rm pa}}$ & ${S_{\rm total}}$ & ${S_{\rm p}}$ & ${\sigma_{\rm rms}}$ & D & ${\xi_{r}}$\\
 &  &  & (mas) & (mas) & (${\degree}$) & (Jy) & (Jy beam$^{{-1}}$) & (mJy beam${^{-1}}$) &  & \\
(1) & (2) & (3) & (4) & (5) & (6) & (7) & (8) & (9) & (10) & (11)\\
\midrule
2013-01-16 & 56308 & K & 6.08 & 3.45 &  78.95 &  3.264 & 3.12 &  2.58 &  355.26 & 0.67 \\
           &       & Q & 3.09 & 1.64 &  78.85 &  2.512 & 2.31 &  5.24 &  183.43 & 0.65 \\
2013-03-28 & 56379 & K & 5.95 & 3.31 & $-84.68$ &  3.192 & 3.03 &  3.01 &  243.38 & 0.68 \\
           &       & Q & 2.95 & 1.73 & $-82.69$ &  2.290 & 2.16 &  5.39 &   98.64 & 0.63 \\
           &       & W & 1.52 & 0.78 & $-89.04$ &  1.457 & 1.15 &  6.22 &   41.89 & 0.84 \\
2013-04-11 & 56393 & K & 5.61 & 3.57 & $-84.14$ &  3.122 & 2.98 &  3.69 &  304.42 & 0.61 \\
           &       & Q & 2.76 & 1.84 & $-86.84$ &  2.345 & 2.19 &  7.64 &  146.80 & 0.69 \\
2013-05-08 & 56420 & K & 6.49 & 3.26 &  78.65 &  3.167 & 2.94 &  3.07 &  200.45 & 0.64 \\
           &       & Q & 3.05 & 1.62 &  77.62 &  2.356 & 2.17 &  6.74 &  130.33 & 0.72 \\
           &       & W & 1.76 & 0.73 &  76.89 &  1.618 & 1.26 &  7.56 &  139.67 & 0.65 \\
           &       & D & 1.15 & 0.52 &  78.44 &  1.322 & 0.92 &  8.91 &   57.91 & 0.65 \\
2013-09-24 & 56559 & K & 5.78 & 3.31 &  88.36 &  2.890 & 2.73 &  5.58 &  161.33 & 0.60 \\
           &       & Q & 2.84 & 1.72 & $-81.86$ &  1.834 & 1.78 &  8.97 &   78.86 & 0.61 \\
2013-11-20 & 56616 & K & 6.07 & 3.20 &  80.73 &  2.920 & 2.80 &  3.10 &  257.46 & 0.60 \\
           &       & Q & 3.06 & 1.59 &  83.13 &  2.259 & 2.07 &  5.84 &  180.43 & 0.58 \\
           &       & W & 1.65 & 0.75 &  80.13 &  1.634 & 1.36 &  6.87 &  145.09 & 0.65 \\
           &       & D & 1.02 & 0.54 &  87.68 &  1.002 & 0.79 & 10.16 &   53.70 & 0.67 \\
2013-12-24 & 56650 & K & 5.82 & 3.28 &  84.49 &  2.967 & 2.67 &  2.31 &  565.44 & 0.56 \\
           &       & Q & 2.92 & 1.65 &  85.51 &  2.147 & 1.99 &  3.85 &  173.50 & 0.67 \\
           &       & W & 1.53 & 0.77 &  82.54 &  1.602 & 1.33 &  5.45 &  139.78 & 0.71 \\
           &       & D & 1.06 & 0.50 &  82.63 &  0.863 & 0.78 &  7.15 &   45.19 & 0.79 \\
2014-01-27 & 56684 & K & 5.84 & 3.39 & $-76.25$ &  2.672 & 2.60 &  3.09 &  295.41 & 0.73 \\
           &       & Q & 2.96 & 1.69 & $-73.44$ &  2.066 & 1.97 &  6.74 &  134.16 & 0.64 \\
2014-02-28 & 56716 & K & 6.41 & 3.17 & $-68.50$ &  2.620 & 2.55 &  3.41 &  267.92 & 0.58 \\
           &       & Q & 3.28 & 1.57 & $-65.36$ &  2.018 & 1.94 &  6.31 &  138.83 & 0.65 \\
           &       & W & 1.89 & 0.70 & $-69.28$ &  1.393 & 1.31 &  9.72 &   56.85 & 0.67 \\
2014-03-22 & 56738 & K & 6.37 & 3.30 & $-65.63$ &  2.533 & 2.46 &  2.09 &  347.31 & 0.63 \\
           &       & Q & 3.38 & 1.57 & $-62.92$ &  1.976 & 1.87 &  5.03 &  217.75 & 0.58 \\
           &       & W & 1.67 & 0.78 & $-67.08$ &  1.492 & 1.32 &  4.83 &   71.59 & 0.81 \\
           &       & D & 1.08 & 0.54 & $-66.86$ &  0.878 & 0.83 &  7.22 &   37.53 & 0.78 \\
2014-04-22 & 56769 & K & 6.29 & 3.35 & $-69.85$ &  2.273 & 2.23 &  2.25 &  135.31 & 0.71 \\
           &       & Q & 3.28 & 1.60 & $-64.58$ &  1.642 & 1.60 &  4.83 &   99.11 & 0.66 \\
           &       & W & 1.50 & 0.78 & $-75.61$ &  1.253 & 1.08 &  7.27 &   35.06 & 0.78 \\
2014-09-01 & 56901 & K & 6.56 & 3.26 & $-67.75$ &  2.661 & 2.49 &  3.97 &  272.19 & 0.67 \\
           &       & Q & 3.36 & 1.61 & $-67.77$ &  2.155 & 2.06 &  4.31 &  344.59 & 0.70 \\
           &       & W & 1.77 & 0.76 & $-67.48$ &  1.563 & 1.43 &  9.95 &  126.25 & 0.85 \\
2014-10-29 & 56959 & K & 6.46 & 3.15 & $-73.62$ &  2.347 & 2.18 &  2.18 &  367.71 & 0.63 \\
           &       & Q & 3.19 & 1.62 & $-76.27$ &  1.689 & 1.55 &  2.88 &  182.22 & 0.68 \\
           &       & W & 1.64 & 0.75 & $-75.79$ &  1.050 & 0.90 &  5.95 &   46.92 & 0.74 \\
2014-11-28 & 56989 & K & 6.33 & 2.88 & $-82.67$ &  2.269 & 1.97 &  6.10 &  159.97 & 0.68 \\
           &       & Q & 3.00 & 1.55 & $-67.76$ &  1.907 & 1.57 &  7.44 &   79.63 & 0.70 \\
2014-12-26 & 57017 & K & 7.06 & 3.12 & $-65.04$ &  2.520 & 2.32 &  1.73 &  342.46 & 0.66 \\
           &       & Q & 3.31 & 1.66 & $-64.55$ &  2.219 & 2.04 &  2.40 &  346.68 & 0.76 \\
           &       & W & 1.82 & 0.76 & $-64.34$ &  1.684 & 1.36 &  4.11 &  124.53 & 0.87 \\
2015-02-24 & 57077 & K & 6.14 & 3.27 & $-71.54$ &  2.745 & 2.59 &  2.22 &  347.84 & 0.61 \\
           &       & Q & 2.88 & 1.76 & $-78.34$ &  2.282 & 2.05 &  3.59 &   90.77 & 0.63 \\
           &       & W & 1.76 & 0.79 & $-67.64$ &  1.567 & 1.29 &  5.85 &   64.64 & 0.68 \\
2015-03-26 & 57107 & K & 6.15 & 3.21 & $-71.00$ &  2.952 & 2.77 &  1.75 &  515.93 & 0.68 \\
           &       & Q & 2.99 & 1.64 & $-69.39$ &  2.376 & 2.25 &  2.82 &  282.80 & 0.76 \\
           &       & W & 1.58 & 0.78 & $-70.34$ &  1.823 & 1.59 &  3.77 &   94.58 & 0.82 \\
           &       & D & 0.99 & 0.56 & $-69.45$ &  1.134 & 1.01 &  4.78 &   59.39 & 0.83 \\
\bottomrule
\end{tabular}
}}
\tabnote{\textbf{Notes.} (1) observing date in yyyy-mm-dd, (2) Modified Julian Date, (3) observing frequency band - `K' for the 22~GHz band, `Q' for the 43~GHz band, `W' for the 86~GHz band, and `D' for the 129~GHz band, (4) major axis of the synthesized beam in mas, (5) minor axis of the synthesized beam in mas, (6) position angle of the major axis in degrees, (7) total CLEANed flux density in Jy, (8) peak flux density in Jy beam$^{-1}$, (9) rms noise in mJy beam$^{-1}$, (10) dynamic range, and (11) image quality factor}
\end{table}
\twocolumn
%%%%%%%%%%%%%%%%%%%%%%%%%%%%%%%%%%%%%%%%%%%%%%%%%%%%%%%%%%%%%%%%%%%%%
%%% TABLE %%%%%%%%%%%%%%%%%%%%%%%%%%%%%%%%%%%%%%%%%%%%%%%%%%%%%%%%%%%
\onecolumn
\begin{table}[hbt!]
\small{\textbf{Table 1.} Continued.}
%\caption{Continued. \label{1}}
\centering
\resizebox{\textwidth}{!}{%
{\scriptsize
\begin{tabular}{lcccccccccc}
\toprule
Date & MJD & Band & ${B_{\rm maj}}$ & ${B_{\rm min}}$ & ${B_{\rm pa}}$ & ${S_{\rm total}}$ & ${S_{\rm p}}$ & ${\sigma_{\rm rms}}$ & D & ${\xi_{r}}$\\
 &  &  & (mas) & (mas) & (${\degree}$) & (Jy) & (Jy beam$^{{-1}}$) & (mJy beam${^{-1}}$) &  & \\
(1) & (2) & (3) & (4) & (5) & (6) & (7) & (8) & (9) & (10) & (11)\\
\midrule
2015-04-30 & 57142 & K & 6.14 & 3.31 & $-72.04$ &  3.036 & 2.76 &  3.12 &  256.33 & 0.65 \\
           &       & Q & 3.07 & 1.72 & $-70.67$ &  2.798 & 2.55 &  4.45 &  227.07 & 0.66 \\
           &       & W & 1.62 & 0.76 & $-71.42$ &  2.214 & 1.68 &  9.31 &   84.62 & 0.71 \\
2015-09-24 & 57289 & K & 7.30 & 2.84 & $-65.90$ &  2.982 & 2.88 &  7.12 &  241.85 & 0.63 \\
           &       & Q & 3.57 & 1.36 & $-74.75$ &  2.455 & 2.36 &  6.43 &  232.76 & 0.59 \\
2015-10-23 & 57318 & K & 6.21 & 3.07 & $-72.29$ &  3.078 & 2.88 &  4.31 &  348.81 & 0.76 \\
           &       & Q & 2.88 & 1.57 & $-86.19$ &  2.612 & 2.44 &  8.66 &  176.45 & 0.70 \\
2015-11-30 & 57356 & K & 6.31 & 3.24 & $-70.00$ &  2.967 & 2.86 &  2.53 &  357.54 & 0.65 \\
           &       & Q & 3.01 & 1.69 & $-74.03$ &  2.477 & 2.33 &  3.79 &  171.26 & 0.61 \\
           &       & W & 1.67 & 0.77 & $-70.13$ &  1.784 & 1.56 &  6.02 &   88.12 & 0.78 \\
           &       & D & 1.12 & 0.50 & $-69.38$ &  1.046 & 0.94 &  6.51 &   36.51 & 0.73 \\
2015-12-28 & 57384 & K & 6.15 & 3.02 & $-89.07$ &  2.667 & 2.48 &  2.52 &  301.21 & 0.63 \\
           &       & Q & 2.90 & 1.59 & $-84.42$ &  2.155 & 2.04 &  3.86 &  253.17 & 0.67 \\
           &       & W & 1.53 & 0.76 & $-83.28$ &  1.417 & 1.28 &  4.89 &  178.45 & 0.70 \\
           &       & D & 1.00 & 0.52 & $-78.13$ &  0.711 & 0.64 &  5.15 &   64.59 & 0.77 \\
2016-02-11 & 57429 & K & 5.99 & 3.36 & $-84.86$ &  2.262 & 2.13 &  4.77 &  210.68 & 0.63 \\
2016-03-01 & 57448 & K & 5.89 & 3.30 & $-74.76$ &  2.363 & 2.20 &  2.41 &  303.51 & 0.67 \\
           &       & Q & 2.81 & 1.76 & $-72.02$ &  1.820 & 1.76 &  3.95 &  264.65 & 0.70 \\
           &       & W & 1.56 & 0.81 & $-71.79$ &  1.219 & 1.09 &  4.75 &   49.68 & 0.80 \\
           &       & D & 1.04 & 0.56 & $-65.16$ &  0.724 & 0.58 &  4.56 &   42.09 & 0.77 \\
2016-04-24 & 57502 & K & 6.25 & 3.36 & $-65.69$ &  2.417 & 2.34 &  2.31 &  203.62 & 0.64 \\
           &       & Q & 3.22 & 1.75 & $-67.23$ &  2.241 & 2.15 &  3.99 &  149.06 & 0.78 \\
           &       & W & 1.73 & 0.82 & $-62.65$ &  1.645 & 1.56 &  6.00 &   78.07 & 0.71 \\
2016-08-23 & 57623 & K & 6.68 & 3.27 & $-59.74$ &  2.788 & 2.67 &  6.19 &  233.38 & 0.66 \\
           &       & Q & 3.40 & 1.62 & $-62.55$ &  2.433 & 2.22 &  6.37 &  182.67 & 0.62 \\
2016-10-18 & 57679 & K & 6.15 & 3.19 & $-74.46$ &  2.896 & 2.74 &  3.48 &  348.69 & 0.65 \\
           &       & Q & 3.02 & 1.61 & $-76.64$ &  2.353 & 2.25 &  6.16 &  140.40 & 0.60 \\
           &       & W & 1.48 & 0.77 & $-79.87$ &  1.647 & 1.45 &  7.42 &   67.08 & 0.79 \\
           &       & D & 0.99 & 0.55 & $-76.71$ &  1.040 & 0.92 & 12.60 &   39.09 & 0.77 \\
2016-11-27 & 57719 & K & 5.72 & 3.57 & $-84.74$ &  2.912 & 2.78 &  3.15 &  353.29 & 0.69 \\
2016-12-28 & 57750 & K & 5.72 & 3.57 & $-84.74$ &  2.921 & 2.79 &  3.15 &  325.44 & 0.65 \\
           &       & Q & 3.02 & 1.56 & $-74.63$ &  2.112 & 2.00 &  3.95 &  274.42 & 0.71 \\
           &       & W & 1.44 & 0.81 & $-75.99$ &  1.402 & 1.24 &  5.35 &   48.52 & 0.74 \\
           &       & D & 0.97 & 0.57 & $-70.57$ &  0.740 & 0.78 &  5.88 &   27.19 & 0.78 \\
2017-03-29 & 57841 & K & 5.68 & 3.30 & $-77.98$ &  2.870 & 2.71 &  2.62 &  472.36 & 0.69 \\
           &       & Q & 3.05 & 1.65 & $-74.83$ &  2.005 & 1.85 &  4.50 &  215.54 & 0.69 \\
           &       & D & 1.18 & 0.46 & $-79.27$ &  0.827 & 0.70 & 12.31 &   51.80 & 0.77 \\
2017-04-19 & 57862 & K & 6.30 & 3.24 & $-75.49$ &  2.690 & 2.55 &  2.75 &  283.88 & 0.69 \\
           &       & Q & 3.03 & 1.51 & $-88.88$ &  1.930 & 1.79 &  5.70 &  152.18 & 0.72 \\
           &       & D & 0.98 & 0.55 &  88.72 &  0.816 & 0.73 &  8.62 &   30.41 & 0.68 \\
2017-06-17 & 57921 & K & 5.96 & 3.03 & $-82.35$ &  2.643 & 2.52 &  4.27 &  223.41 & 0.63 \\
           &       & Q & 3.77 & 1.41 & $-64.62$ &  1.788 & 1.72 &  4.65 &  143.82 & 0.71 \\
           &       & W & 1.76 & 0.73 & $-66.95$ &  1.284 & 1.13 &  9.71 &   36.08 & 0.64 \\
           &       & D & 1.08 & 0.51 & $-64.85$ &  0.790 & 0.77 & 21.26 &   23.96 & 0.66 \\
2017-09-19 & 58015 & K & 6.32 & 3.00 & $-69.39$ &  2.778 & 2.55 &  3.37 &  188.85 & 0.60 \\
           &       & Q & 3.38 & 1.39 & $-79.23$ &  1.566 & 1.49 &  7.41 &   89.63 & 0.62 \\
           &       & W & 1.52 & 0.76 & $-81.56$ &  1.166 & 1.14 &  9.80 &   56.68 & 0.71 \\
2017-10-21 & 58047 & K & 6.14 & 3.00 & $-81.30$ &  2.359 & 2.24 &  4.53 &  186.88 & 0.67 \\
           &       & W & 1.63 & 0.71 & $-84.67$ &  1.000 & 0.93 &  6.53 &   84.85 & 0.74 \\
           &       & D & 1.03 & 0.50 & $-85.35$ &  0.507 & 0.51 &  8.18 &   28.81 & 0.77 \\
2017-11-05 & 58062 & K & 6.84 & 2.76 & $-86.60$ &  2.188 & 2.03 &  3.97 &  250.39 & 0.65 \\
           &       & Q & 3.62 & 1.59 & $-65.32$ &  1.546 & 1.48 &  2.85 &  114.79 & 0.70 \\
           &       & W & 1.77 & 0.82 & $-62.37$ &  1.116 & 1.09 &  4.89 &   33.24 & 0.69 \\
           &       & D & 1.11 & 0.57 & $-61.17$ &  0.766 & 0.78 &  6.84 &   23.06 & 0.73 \\
\bottomrule
\end{tabular}
}}
\end{table}
\twocolumn
%%%%%%%%%%%%%%%%%%%%%%%%%%%%%%%%%%%%%%%%%%%%%%%%%%%%%%%%%%%%%%%%%%%%%
%%% TABLE %%%%%%%%%%%%%%%%%%%%%%%%%%%%%%%%%%%%%%%%%%%%%%%%%%%%%%%%%%%
\onecolumn
\begin{table}[hbt!]
\small{\textbf{Table 1.} Continued.}
%\caption{Continued.}
\centering
\resizebox{\textwidth}{!}{%
{\scriptsize
\begin{tabular}{lcccccccccc}
\toprule
Date & MJD & Band & ${B_{\rm maj}}$ & ${B_{\rm min}}$ & ${B_{\rm pa}}$ & ${S_{\rm total}}$ & ${S_{\rm p}}$ & ${\sigma_{\rm rms}}$ & D & ${\xi_{r}}$\\
 &  &  & (mas) & (mas) & (${\degree}$) & (Jy) & (Jy beam$^{{-1}}$) & (mJy beam${^{-1}}$) &  & \\
(1) & (2) & (3) & (4) & (5) & (6) & (7) & (8) & (9) & (10) & (11)\\
\midrule
2017-11-22 & 58079 & K & 7.83 & 3.07 & $-61.81$ &  2.194 & 2.08 &  2.17 &  345.13 & 0.67 \\
           &       & Q & 3.19 & 1.73 &  84.27 &  1.553 & 1.43 &  4.76 &  173.69 & 0.75 \\
           &       & D & 0.99 & 0.58 & $-82.49$ &  0.611 & 0.55 &  6.37 &   30.37 & 0.70 \\
2018-01-13 & 58131 & K & 6.26 & 3.48 &  83.71 &  2.110 & 1.97 &  2.86 &  245.87 & 0.63 \\
           &       & Q & 3.12 & 1.50 & $-84.81$ &  1.801 & 1.62 &  3.41 &  257.83 & 0.69 \\
           &       & W & 1.48 & 0.81 & $-75.02$ &  1.191 & 1.14 &  4.15 &   51.16 & 0.82 \\
           &       & D & 0.99 & 0.56 & $-72.98$ &  0.657 & 0.67 &  4.54 &   29.38 & 0.86 \\
2018-02-19 & 58168 & K & 6.04 & 3.32 & $-67.28$ &  2.296 & 2.17 &  2.23 &  175.73 & 0.63 \\
           &       & Q & 3.21 & 1.62 & $-75.67$ &  1.921 & 1.80 &  3.41 &  241.54 & 0.72 \\
           &       & W & 1.61 & 0.84 & $-76.28$ &  1.402 & 1.38 &  5.94 &   82.66 & 0.78 \\
           &       & D & 0.98 & 0.61 & $-88.55$ &  0.908 & 0.90 &  7.93 &  114.84 & 0.71 \\
2018-03-26 & 58203 & K & 6.20 & 3.47 & $-67.78$ &  2.524 & 2.33 &  3.14 &  301.39 & 0.74 \\
           &       & Q & 3.08 & 1.53 &  85.02 &  2.043 & 1.87 &  3.47 &  193.26 & 0.80 \\
           &       & W & 1.50 & 0.78 &  86.91 &  1.612 & 1.38 &  4.88 &   48.14 & 0.79 \\
           &       & D & 0.95 & 0.58 & $-78.73$ &  1.033 & 0.84 &  4.57 &   36.56 & 0.84 \\
2018-04-19 & 58227 & K & 5.89 & 3.49 & $-71.84$ &  2.804 & 2.50 &  2.14 &  203.12 & 0.70 \\
           &       & Q & 3.28 & 1.65 & $-70.75$ &  2.164 & 1.93 &  3.43 &  170.21 & 0.67 \\
           &       & W & 1.67 & 0.79 & $-63.70$ &  1.841 & 1.55 &  5.44 &   61.36 & 0.68 \\
           &       & D & 1.27 & 0.55 & $-53.42$ &  1.240 & 1.00 &  8.08 &   35.24 & 0.66 \\
2018-05-16 & 58254 & K & 6.16 & 3.20 & $-69.54$ &  2.674 & 2.51 &  2.02 &  282.23 & 0.63 \\
           &       & W & 1.68 & 0.74 & $-65.13$ &  1.917 & 1.76 &  9.84 &   84.40 & 0.66 \\
2018-05-31 & 58269 & K & 5.69 & 3.20 & $-76.07$ &  2.777 & 2.55 &  3.33 &  147.11 & 0.63 \\
           &       & Q & 3.01 & 1.74 &  69.41 &  2.728 & 2.58 &  6.06 &   94.69 & 0.63 \\
2018-12-14 & 58466 & K & 5.67 & 3.33 & $-74.41$ &  2.897 & 2.68 &  3.83 &  245.74 & 0.60 \\
2019-01-26 & 58509 & K & 5.91 & 3.47 & $-59.95$ &  3.464 & 3.27 &  3.95 &  275.94 & 0.69 \\
           &       & Q & 3.30 & 1.83 & $-87.35$ &  2.297 & 2.08 &  3.95 &  175.24 & 0.65 \\
           &       & W & 1.47 & 0.86 & $-85.06$ &  1.652 & 1.30 &  7.29 &   42.70 & 0.61 \\
2019-02-15 & 58529 & K & 6.33 & 3.70 & $-74.42$ &  3.351 & 3.04 &  3.41 &  132.54 & 0.64 \\
           &       & Q & 2.88 & 1.72 &  79.49 &  2.083 & 1.94 &  6.99 &   86.97 & 0.59 \\
2019-04-20 & 58593 & K & 5.85 & 3.63 & $-64.47$ &  2.998 & 2.86 &  4.34 &  151.18 & 0.68 \\
           &       & Q & 2.88 & 1.65 & $-75.07$ &  1.911 & 1.72 &  4.45 &  104.15 & 0.66 \\
           &       & W & 1.40 & 0.82 & $-81.55$ &  1.070 & 0.98 &  7.21 &   47.23 & 0.61 \\
           &       & D & 0.93 & 0.58 & $-81.59$ &  0.684 & 0.62 & 12.83 &   41.45 & 0.64 \\
2019-06-03 & 58637 & K & 5.80 & 3.43 & $-69.51$ &  2.678 & 2.48 &  2.92 &  271.55 & 0.65 \\
           &       & Q & 3.93 & 1.54 & $-62.36$ &  2.077 & 1.82 &  3.15 &  149.71 & 0.61 \\
           &       & W & 1.71 & 0.83 & $-60.69$ &  1.231 & 1.09 &  5.25 &   67.08 & 0.76 \\
           &       & D & 1.13 & 0.57 & $-59.85$ &  0.740 & 0.64 &  6.88 &   31.10 & 0.77 \\
2019-06-24 & 58658 & K & 7.43 & 3.33 & $-59.93$ &  2.600 & 2.47 &  2.73 &  148.63 & 0.67 \\
           &       & Q & 3.07 & 1.59 & $-71.01$ &  1.844 & 1.69 &  2.81 &  209.22 & 0.75 \\
           &       & W & 1.53 & 0.84 & $-67.70$ &  0.971 & 0.92 &  6.04 &   64.12 & 0.77 \\
2020-01-08 & 58856 & K & 5.76 & 3.40 & $-73.51$ &  2.623 & 2.38 &  2.48 &  313.07 & 0.80 \\
           &       & Q & 3.25 & 1.61 & $-72.15$ &  1.291 & 1.19 &  2.78 &  165.17 & 0.69 \\
           &       & W & 1.59 & 0.83 & $-68.5$6 &  0.692 & 0.62 &  4.98 &   55.74 & 0.72 \\
           &       & D & 1.17 & 0.55 & $-68.54$ &  0.324 & 0.31 & 10.62 &   21.81 & 0.71 \\
2020-02-13 & 58892 & K & 6.32 & 3.39 & $-67.53$ &  1.983 & 1.84 &  2.19 &  187.77 & 0.77 \\
           &       & Q & 3.23 & 1.62 & $-70.57$ &  1.179 & 1.06 &  2.79 &  123.16 & 0.67 \\
           &       & W & 1.65 & 0.82 & $-69.36$ &  0.645 & 0.51 &  4.13 &   39.81 & 0.71 \\
2020-03-13 & 58921 & K & 6.37 & 3.43 & $-68.18$ &  1.809 & 1.67 &  2.07 &  155.09 & 0.74 \\
           &       & Q & 3.01 & 1.59 & $-71.67$ &  1.097 & 1.01 &  3.12 &  107.11 & 0.77 \\
           &       & W & 1.55 & 0.82 & $-73.18$ &  0.518 & 0.42 &  3.89 &   27.25 & 0.73 \\
\bottomrule
\end{tabular}
}}
\end{table}
\twocolumn
%%%%%%%%%%%%%%%%%%%%%%%%%%%%%%%%%%%%%%%%%%%%%%%%%%%%%%%%%%%%%%%%%%%%%

\subsection{Fermi-LAT Data}
The ${\gamma}$-ray data of 4C~+28.07 are obtained from the Fermi Large Array Telescope~(LAT) Light Curve Repository~\citep[LCR;][]{abdollahi2023fermi}~\footnote{\url{https://fermi.gsfc.nasa.gov/ssc/data/access/lat/LightCurveRepository/}}. The Fermi-LAT is an instrument of the Fermi Gamma-ray Space Telescope mission, covering an energy range of 20~MeV to 300~GeV~\citep{atwood2009large}. The Fermi-LAT LCR analysis, covering the energy range from 100~MeV to 100~GeV, is conducted via an unbinned likelihood analysis to obtain the light curves using the standard LAT Fermitools~\footnote{\url{https://fermi.gsfc.nasa.gov/ssc/data/analysis/software/}}~(version 1.0.5). Following~\citet{abdollahi2023fermi}, there are several points to note with the LCR~\citep[see][for more details]{abdollahi2023fermi}. Initially, in order to utilize higher-level analysis methods~(e.g., multifrequency cross-correlation), the end user should regard non-convergent data as suspicious. Secondly, because the LCR analysis was conducted in real time, the data from the LCR was not validated by the LAT Collaboration. Therefore, we conducted an additional investigation, removed outliers from the data, and then proceeded with the data analysis. We selected ${\gamma}$-ray data of 4C~+28.07 with a cadence of 7-days.

\section{Results}\label{sec:4}

\subsection{Multiwavelength Light Curves}
\begin{figure*}[hbt!]
\centering
\includegraphics[width=\textwidth]{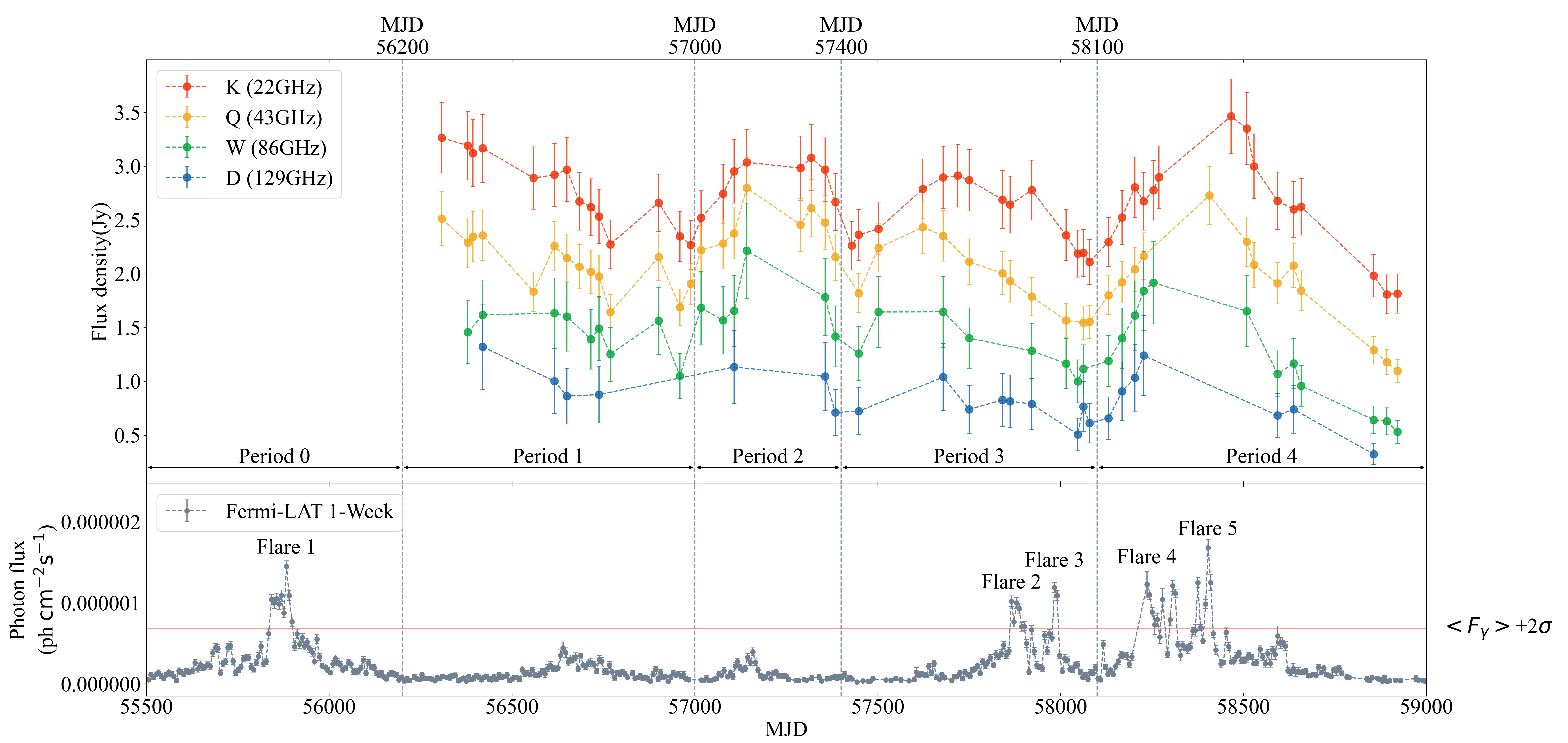}
\caption{Light curves of 4C~+28.07. The total CLEANed flux densities at 22, 43, 86, and 129~GHz are plotted in red, yellow, green, and blue, respectively. In the bottom plot, the Fermi-LAT data are in gray, and the orange-colored solid line is a threshold~(the sum of the mean value and twice the standard deviation, ${<F_{\gamma}>+2\sigma}$) for ${\gamma}$-ray flares. \label{fig:lc}}
\end{figure*}

The multiwavelength light curves are shown in Figure~\ref{fig:lc}, where the iMOGABA data~(22-129~GHz) are plotted using the total CLEANed flux density obtained from the CLEAN script, and the Fermi-LAT data are plotted using 7-day binned data.
%%% FIGURE %%%%%%%%%%%%%%%%%%%%%%%%%%%%%%%%%%%%%%%%%%%%%%%%%%%%%%%%%%
%%%%%%%%%%%%%%%%%%%%%%%%%%%%%%%%%%%%%%%%%%%%%%%%%%%%%%%%%%%%%%%%%%%%%

\subsubsection{22-129~GHz Light Curves}
As shown in Figure~\ref{fig:lc}, 4C~+28.07 has various flux densities of radio and ${\gamma}$-ray, in particular, yielding several flares in the ${\gamma}$-ray. The multifrequency radio light curves are described in this Section. The multifrequency total CLEANed flux densities~(${S_{\rm total}}$) are in the corresponding ranges of 1.81-3.46~Jy, 1.10-2.80~Jy, 0.53-2.21~Jy, 0.32-1.32~Jy at 22, 43, 86 and 129~GHz. The uncertainties are 10\%, 10\%, 20\%, and 30\% of ${S_{\rm total}}$ at 22, 43, 86 and 129~GHz, respectively. There are less number of data at 129~GHz compared to those at 22-86~GHz mainly due to no fringe detection at 129~GHz (i.e., poor weather) or no recording at 129~GHz with three-frequency observation.

For further analysis, the light curves are divided into five periods~(periods~0-4) based on the local minima of the 22~GHz light curves with having the periods including the active $\gamma$-ray periods and the dividing epochs in the quiet $\gamma$-ray periods: the period of MJD~55500-56200 for period~0, the period of MJD~56200-57000 for period~1, the period of MJD~57000-57400 for period~2, the period of MJD~57400-58100 for period~3, and the period of MJD~58100-59000 for period~4. The reason why we introduced the period~0 is to understand the decreasing trends in the radio light curve with a potential relation with the $\gamma$-ray activity in the period~0~(see Section \ref{sec:multifrequency variability characteristics}).

In period~1, clear decreases by about 1~Jy of the 22 and 43~GHz light curves are found. In period~2, the radio flux densities ${S_{\rm{total}}}$ increases at 22-86~GHz, and in period~3, the 22~GHz data show a gentle flux change overall, but the 43~GHz and 86~GHz data show the peak of ${S_{\rm{total}}}$ in the first half of period~3. In period~4, there are very few data points at 22-129~GHz, but unlike period~3, the peaks may appear in a similar period.

\subsubsection{\texorpdfstring{${\gamma}$}--ray Light Curve}
We obtained Fermi-LAT data for the period of MJD~55500 to MJD~59000. We have defined the flares in the gamma-ray light curve with a threshold of $<F_{\gamma}>+2\sigma$ ($<F_{\gamma}>$ is a mean of the light curve and $\sigma$ a standard deviation of the light curve). And we defined 5 flares in periods~0, 3, and 4, applying the threshold, i.e., peaks above the threshold are selected as flares. Two weak ${\gamma}$-ray enhancements are also seen in periods~1, and 2, although they are under the selection criteria. In this paper, we used multifrequency radio VLBI data in periods~1-4 to study possible multifrequency correlations of the flares.

\subsection{Fractional Variability Amplitude}
\begin{figure}[hbt!]
\centering
\includegraphics[width=77.5mm]{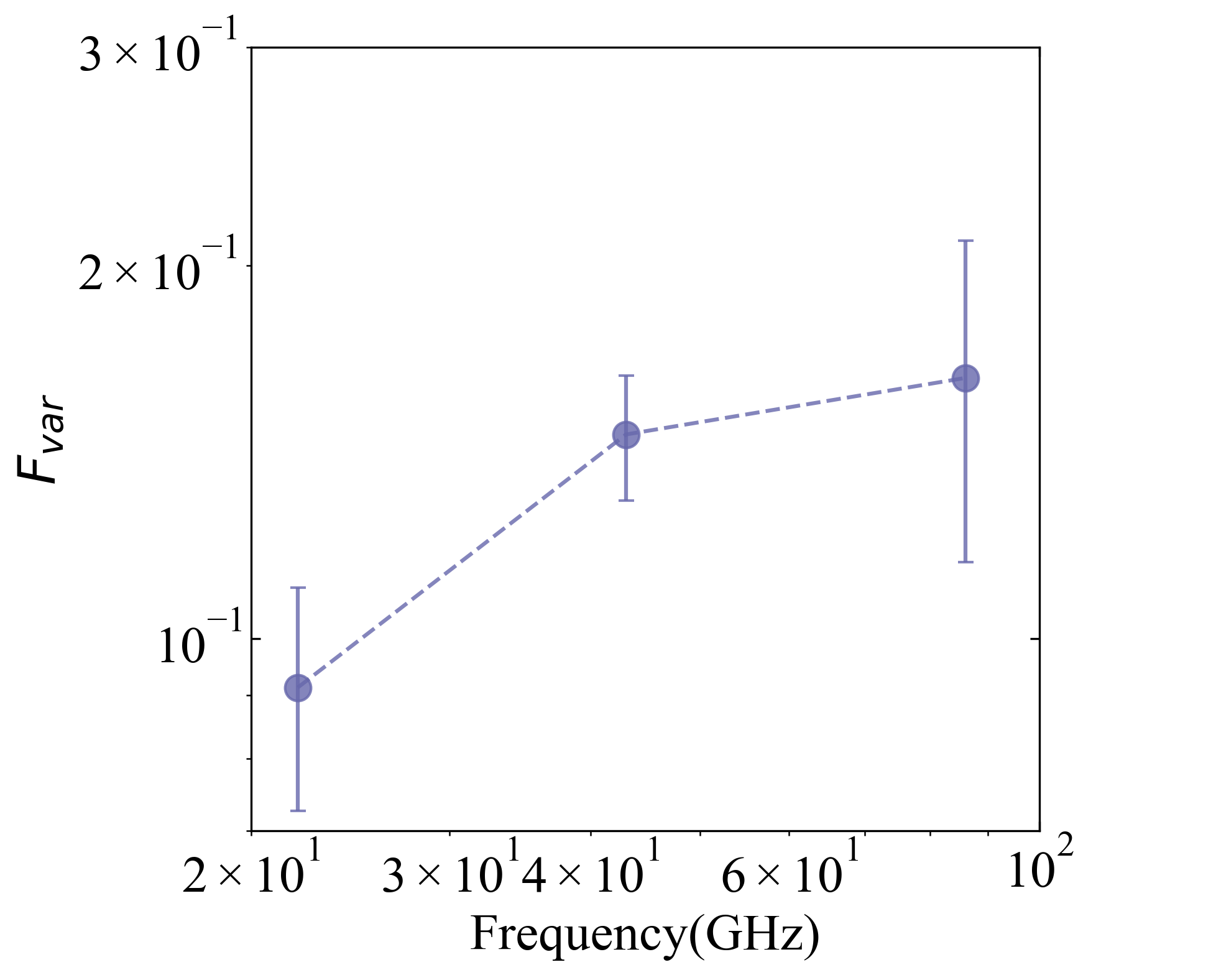}
\caption{Fractional variability amplitudes of 4C~+28.07 obtained from the KVN multifrequency light curves at 22-129~GHz\label{fig:frac_var}}
\end{figure}

The multifrequency flux density variability outcomes are compared by computing the fractional variability amplitude~(${F_{\rm{var}}}$)~\citep{chidiac2016exploring, lee2020interferometric}, which is defined as
\begin{equation}
    F_{\rm{var}} = \sqrt{\frac{V-\overline{\sigma_{\rm{err}}^{2}}}{{\overline{S}}^2}},
\end{equation}
where $V$ is the variance of the flux density, $\overline{\sigma_{\rm{err}}^{2}}$ is the mean squared error, and ${\overline{S}}$ is the mean flux density. The uncertainty of the fractional variability is estimated by
\begin{equation}
    \rm{err}(\textit F_{\rm{var}}) = \sqrt{\left(\sqrt{\frac{1}{2 \textit N}}\frac{\overline{\sigma_{\rm{err}}^{2}}}{{\overline{ \textit S}}^{2}F_{\rm{var}}}\right)^{2} + \left(\sqrt{\frac{\overline{\sigma_{\rm{err}}^{2}}}{\textit N}} \frac{1}{\overline{\textit S}}\right)^2},
\end{equation}
where $N$ is the amount of data for each band. The detailed results of $F_{\rm{var}}$ are shown in Table~\ref{tab:frac_var}. The uncertainty of 129~GHz exceeded the variance of 129~GHz; therefore, ${F_{\rm{var}}}$ could not be obtained. As shown in Figure~\ref{fig:frac_var}, ${F_{\rm{var}}}$ increases as a function of the frequency, indicating that the variability amplitude of the source becomes larger at a higher frequency.

%%% TABLE %%%%%%%%%%%%%%%%%%%%%%%%%%%%%%%%%%%%%%%%%%%%%%%%%%%%%%%%%%%
%appendix
\begin{table}[b!]
\caption{\label{tab:frac_var} Fractional variability of ${S_{\rm{total}}}$.}
\centering
\begin{tabular}{cc}
\toprule
Frequency & ${F_{\rm{var}}}$\\
(GHz) & \\
\midrule
22  & 0.09$\pm$0.02\\
43  & 0.15$\pm$0.02\\
86  & 0.16$\pm$0.05\\
\bottomrule
\end{tabular}
\end{table}
%%%%%%%%%%%%%%%%%%%%%%%%%%%%%%%%%%%%%%%%%%%%%%%%%%%%%%%%%%%%%%%%%%%%%

%%% FIGURE %%%%%%%%%%%%%%%%%%%%%%%%%%%%%%%%%%%%%%%%%%%%%%%%%%%%%%%%%%
%%%%%%%%%%%%%%%%%%%%%%%%%%%%%%%%%%%%%%%%%%%%%%%%%%%%%%%%%%%%%%%%%%%%%

\subsection{Variable Flux} 
\begin{figure}[hbt!]
\centering
\includegraphics[width=70mm]{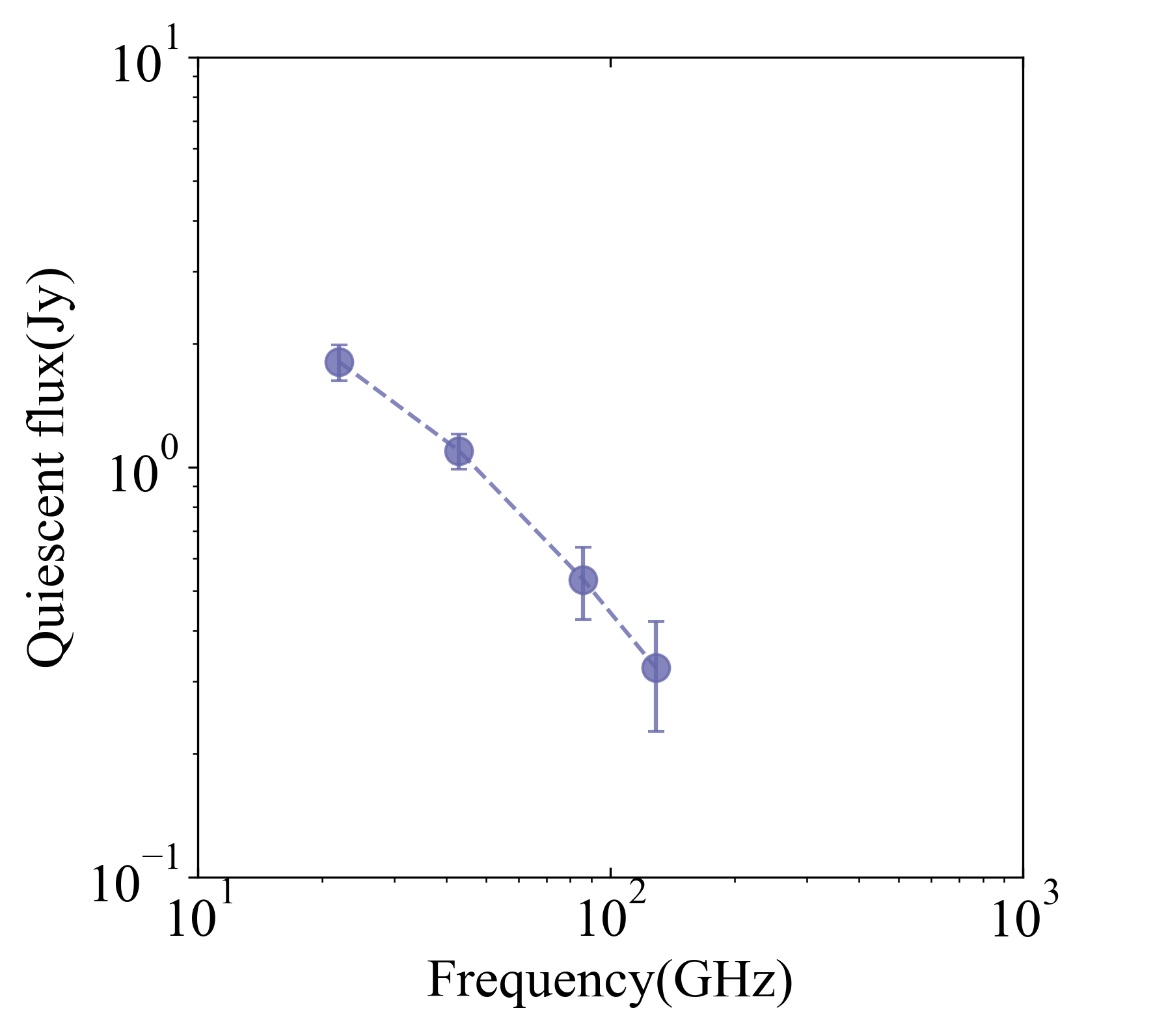}
\caption{Spectrum of the quiescent flux densities of 4C~+28.07 obtained from the KVN light curves at 22-129~GHz \label{fig:flux_qui}}
\end{figure}

\begin{figure*}[hbt!]
\centering
\includegraphics[width=\textwidth]{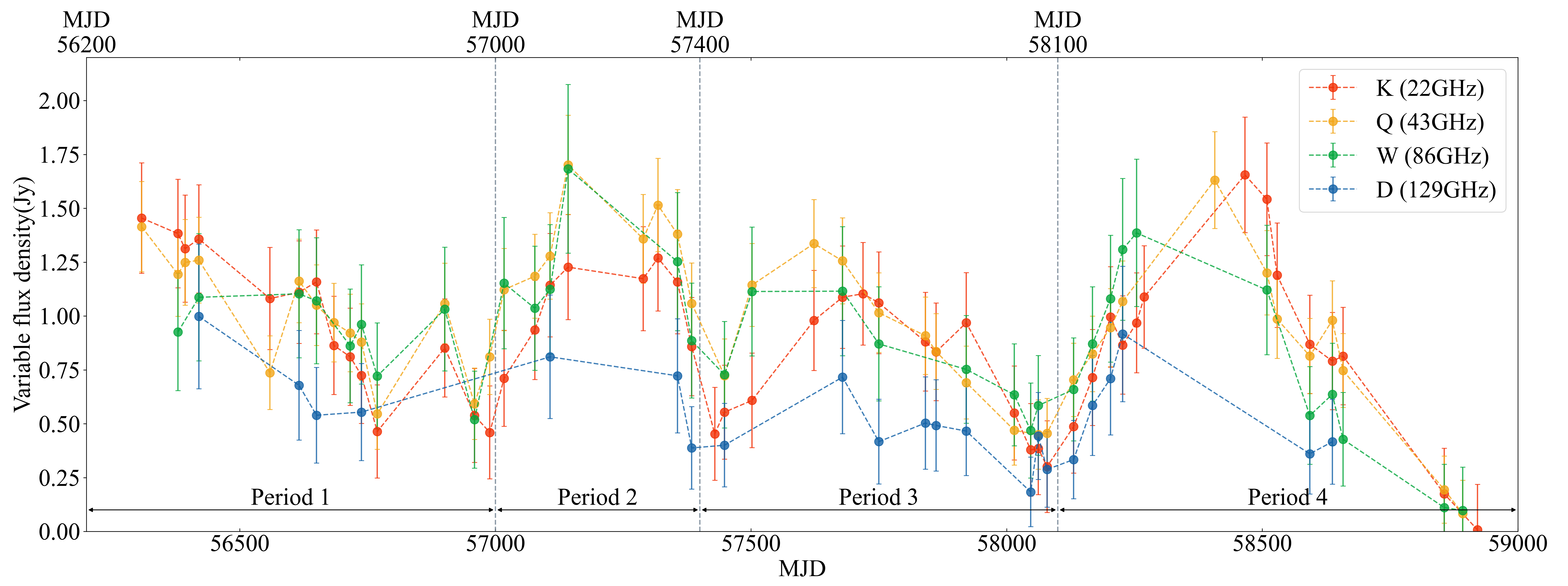}
\caption{
The light curves of the variable flux densities. Red, yellow, green, and blue dots are the variable flux densities at 22, 43, 86, and 129~GHz, respectively.
\label{fig:lc_cq}}
\end{figure*}

The multifrequency radio light curves in this work show that the minimum flux density is not zero, implying that even on the mas scale, constant radiation is observed from the target source in this period~(i.e., a quiescent emission region). In order to study the physical properties~(e.g., spectral properties) of the variable emission region in the source, we determined the quiescent emission spectrum of the source by measuring the minimum flux densities ${S_{\rm{q}}}$ at 22-129~GHz, as summarized in Table~\ref{tab:quiescent_flux} and as shown in Figure~\ref{fig:flux_qui}. We found that the spectrum of the quiescent flux densities is optically thin, indicating that the quiescent emission may come from or is a part of optically thin regions in the source. The quiescent spectrum as determined is subtracted from the multifrequency light curves, yielding the variable flux density ${S_{\rm{var}}}$, as shown in Figure~\ref{fig:lc_cq}. We found that the variable flux densities appear to have a flat spectrum, details of which will be described in the following section.

%%% FIGURE %%%%%%%%%%%%%%%%%%%%%%%%%%%%%%%%%%%%%%%%%%%%%%%%%%%%%%%%%%
%%%%%%%%%%%%%%%%%%%%%%%%%%%%%%%%%%%%%%%%%%%%%%%%%%%%%%%%%%%%%%%%%%%%%
%%% TABLE %%%%%%%%%%%%%%%%%%%%%%%%%%%%%%%%%%%%%%%%%%%%%%%%%%%%%%%%%%%
\begin{table}[hbt!]
\caption{\label{tab:quiescent_flux} Quiescent fluxes of 22-129~GHz data.}
\centering
\begin{tabular}{lccc}
\toprule
Date & MJD & Frequency & ${S_{\rm{q}}}$\\
 &  & (GHz) & (Jy)\\
\midrule
2020-02-13 & 58892 & 22  & 1.809$\pm$0.181\\
2020-03-13 & 58921 & 43  & 1.097$\pm$0.110\\
2020-03-13 & 58921 & 86  & 0.532$\pm$0.106\\
2020-01-08 & 58856 & 129 & 0.324$\pm$0.097\\
\bottomrule
\end{tabular}
\end{table}
%%%%%%%%%%%%%%%%%%%%%%%%%%%%%%%%%%%%%%%%%%%%%%%%%%%%%%%%%%%%%%%%%%%%%
%%% FIGURE %%%%%%%%%%%%%%%%%%%%%%%%%%%%%%%%%%%%%%%%%%%%%%%%%%%%%%%%%%
%%%%%%%%%%%%%%%%%%%%%%%%%%%%%%%%%%%%%%%%%%%%%%%%%%%%%%%%%%%%%%%%%%%%%

\subsection{Spectral Index}
In order to investigate the spectral properties of the multifrequency simultaneous observation data, we fitted the spectra using two power-law functions following \citet{lee2016interferometric}. The first is the single power-law function is given by
\begin{equation}\label{eq:single_pl}
    S_{\nu} \propto \nu^{\alpha},
\end{equation}
where ${\nu}$ is the observing frequency, ${\alpha}$ is the spectral index, and ${S_{\nu}}$ is the flux density at ${\nu}$. The second is the curved power-law function, expressed as
\begin{equation}\label{eq:curved_pl}
    S_{\nu} = S_{r} \left( \frac{\nu}{\nu_{r}}\right)^{\alpha + b \ln{(\nu/\nu_{r})}},
\end{equation}
where ${S_{\rm r}}$ is the flux density at the reference frequency~(${\nu_{\rm r}}$), ${\alpha}$ is the spectral index, and ${b}$ is a constant.

%%% FIGURE %%%%%%%%%%%%%%%%%%%%%%%%%%%%%%%%%%%%%%%%%%%%%%%%%%%%%%%%%%
\begin{figure*}[p]
\centering
\includegraphics[width=\textwidth]{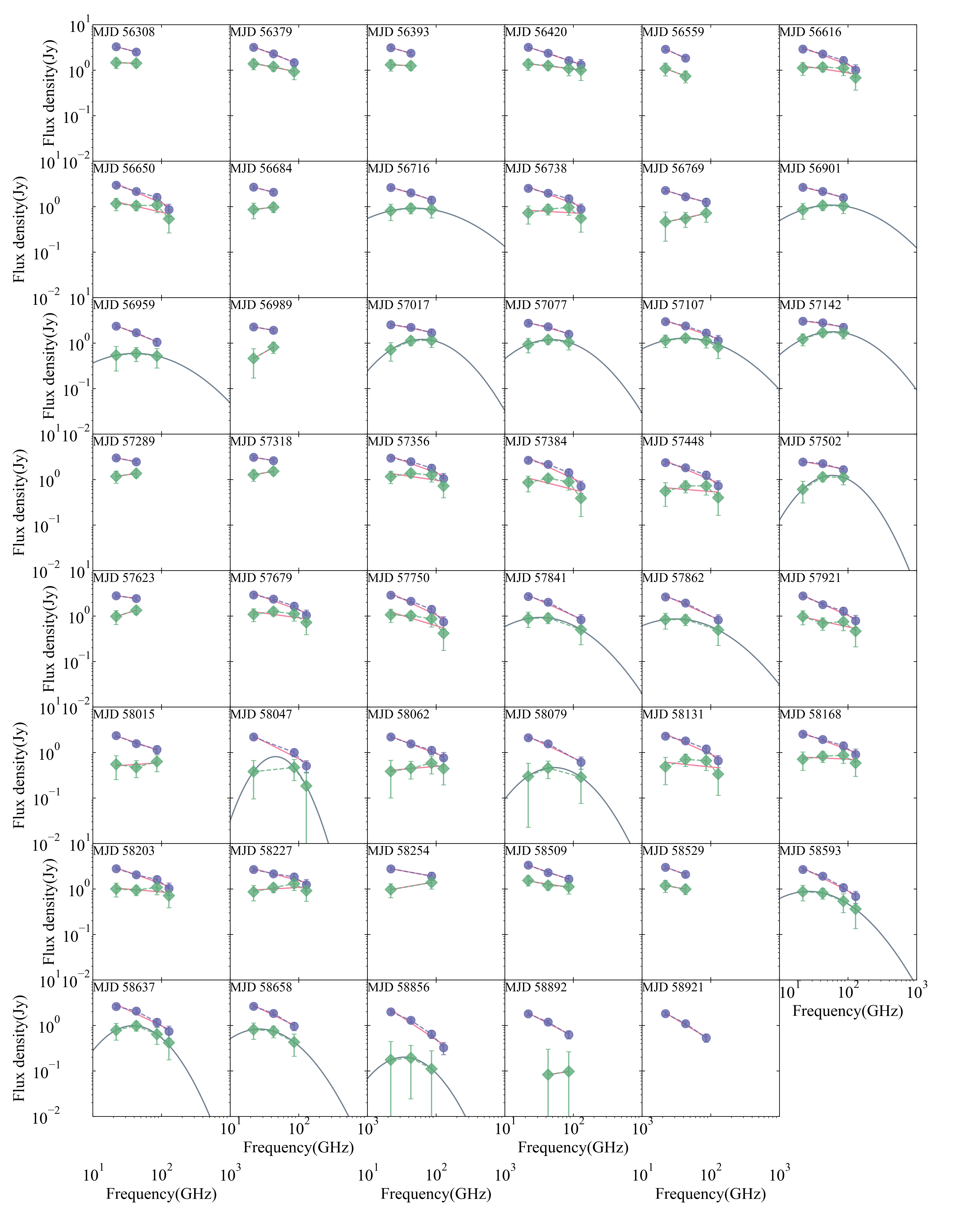}
\caption{Spectra of 4C~+28.07. One graph corresponds to one epoch, the lavender circles are drawn in the total CLEANed flux densities, and the green diamonds are drawn in the variable flux densities. The pink line is the result obtained using the single power-law, and the gray curve is the result obtained using the curved power-law. The dashed line connecting each data point represents the spectrum between adjacent frequencies.\label{fig:pl}}
\end{figure*}
%%%%%%%%%%%%%%%%%%%%%%%%%%%%%%%%%%%%%%%%%%%%%%%%%%%%%%%%%%%%%%%%%%%%%

The fitting results are shown in Figure~\ref{fig:pl}, where ${S_{\rm total}}$ are in lavender and ${S_{\rm var}}$ are in green. Here, we estimated the spectral indices~(e.g., 22-43~GHz, 43-86~GHz, and 86-129~GHz) between adjacent frequencies, as shown by the dashed lines, and we used a single power-law function to fit the spectral indices of the entire frequency used in each epoch, as shown by the solid lines. We found that the spectra of ${S_{\rm total}}$ are optically thin at 22~GHz to 129~GHz. Detailed results for the spectral indices of ${S_{\rm total}}$ are summarized in Table~\ref{tab:spectral_indices_CLEAN}, which lists the spectral indices of adjacent frequencies~(${\alpha_{\rm{22-43}}}$, ${\alpha_{\rm{43-86}}}$, and ${\alpha_{\rm{86-129}}}$) and the spectral indices of the whole frequencies~(${\alpha_{\rm{whole}}}$) for each epoch, and the results of the reduced chi-square method~(${\chi_{\rm{reduced}}}$) for ${\alpha_{\rm{whole}}}$. The uncertainties of the spectral indices in Table~\ref{tab:spectral_indices_CLEAN} were calculated using the error propagation method.

%%% TABLE %%%%%%%%%%%%%%%%%%%%%%%%%%%%%%%%%%%%%%%%%%%%%%%%%%%%%%%%%%%
%appendix
\begin{table*}[p]
\caption{\label{tab:spectral_indices_CLEAN} Spectral indices for the total CLEANed flux densities.}
\centering
\resizebox{\textwidth}{!}{%
{\tiny
\begin{tabular}{lcccccc}
\toprule
 Date & MJD & ${\alpha_{\rm{22-43}}}$ & ${\alpha_{\rm{43-86}}}$ & ${\alpha_{\rm{86-129}}}$ & ${\alpha_{\rm{whole}}^{*}}$ & ${\chi_{\rm{reduced}}}$ \\
(1) & (2) & (3) & (4) & (5) & (6) & (7) \\
\midrule
2013-01-16 & 56308 & ${-0.39\pm0.21}$ &                  &                  & ${-0.39\pm0.21}$ &          \\
2013-03-28 & 56379 & ${-0.50\pm0.21}$ & ${-0.65\pm0.32}$ &                  & ${-0.58\pm0.16}$ & 2.11e-02 \\
2013-04-11 & 56393 & ${-0.43\pm0.21}$ &                  &                  & ${-0.43\pm0.21}$ &          \\
2013-05-08 & 56420 & ${-0.44\pm0.21}$ & ${-0.54\pm0.32}$ & ${-0.50\pm0.89}$ & ${-0.50\pm0.18}$ & 5.31e-03 \\
2013-09-24 & 56559 & ${-0.68\pm0.21}$ &                  &                  & ${-0.68\pm0.21}$ &          \\
2013-11-20 & 56616 & ${-0.38\pm0.21}$ & ${-0.47\pm0.32}$ & ${-1.21\pm0.89}$ & ${-0.57\pm0.18}$ & 1.16e-01 \\
2013-12-24 & 56650 & ${-0.48\pm0.21}$ & ${-0.42\pm0.32}$ & ${-1.53\pm0.89}$ & ${-0.64\pm0.18}$ & 1.31e-01 \\
2014-01-27 & 56684 & ${-0.38\pm0.21}$ &                  &                  & ${-0.38\pm0.21}$ &          \\
2014-02-28 & 56716 & ${-0.39\pm0.21}$ & ${-0.54\pm0.32}$ &                  & ${-0.46\pm0.16}$ & 2.71e-02 \\
2014-03-22 & 56738 & ${-0.37\pm0.21}$ & ${-0.40\pm0.32}$ & ${-1.31\pm0.89}$ & ${-0.56\pm0.18}$ & 1.52e-01 \\
2014-04-22 & 56769 & ${-0.48\pm0.21}$ & ${-0.39\pm0.32}$ &                  & ${-0.44\pm0.16}$ & 1.16e-02 \\
2014-09-01 & 56901 & ${-0.31\pm0.21}$ & ${-0.46\pm0.32}$ &                  & ${-0.39\pm0.16}$ & 3.90e-02 \\
2014-10-29 & 56959 & ${-0.49\pm0.21}$ & ${-0.69\pm0.32}$ &                  & ${-0.59\pm0.16}$ & 3.11e-02 \\
2014-11-28 & 56989 & ${-0.26\pm0.21}$ &                  &                  & ${-0.26\pm0.21}$ &          \\\addlinespace
2014-12-26 & 57017 & ${-0.19\pm0.21}$ & ${-0.40\pm0.32}$ &                  & ${-0.30\pm0.16}$ & 1.34e-01 \\
2015-02-24 & 57077 & ${-0.28\pm0.21}$ & ${-0.54\pm0.32}$ &                  & ${-0.41\pm0.16}$ & 1.17e-01 \\
2015-03-26 & 57107 & ${-0.32\pm0.21}$ & ${-0.52\pm0.32}$ & ${-0.93\pm0.89}$ & ${-0.53\pm0.18}$ & 9.49e-02 \\
2015-04-30 & 57142 & ${-0.12\pm0.21}$ & ${-0.34\pm0.32}$ &                  & ${-0.23\pm0.16}$ & 2.25e-01 \\
2015-09-24 & 57289 & ${-0.29\pm0.21}$ &                  &                  & ${-0.29\pm0.21}$ &          \\
2015-10-23 & 57318 & ${-0.25\pm0.21}$ &                  &                  & ${-0.25\pm0.21}$ &          \\
2015-11-30 & 57356 & ${-0.27\pm0.21}$ & ${-0.47\pm0.32}$ & ${-1.32\pm0.89}$ & ${-0.56\pm0.18}$ & 2.24e-01 \\
2015-12-28 & 57384 & ${-0.32\pm0.21}$ & ${-0.60\pm0.32}$ & ${-1.70\pm0.89}$ & ${-0.70\pm0.18}$ & 2.68e-01 \\\addlinespace
2016-03-01 & 57448 & ${-0.39\pm0.21}$ & ${-0.53\pm0.32}$ & ${-1.37\pm0.89}$ & ${-0.63\pm0.18}$ & 1.41e-01 \\
2016-04-24 & 57502 & ${-0.11\pm0.21}$ & ${-0.45\pm0.32}$ &                  & ${-0.28\pm0.16}$ & 3.61e-01 \\
2016-08-23 & 57623 & ${-0.20\pm0.21}$ &                  &                  & ${-0.20\pm0.21}$ &          \\
2016-10-18 & 57679 & ${-0.31\pm0.21}$ & ${-0.51\pm0.32}$ & ${-1.13\pm0.89}$ & ${-0.56\pm0.18}$ & 1.45e-01 \\
2016-12-28 & 57750 & ${-0.46\pm0.21}$ & ${-0.59\pm0.32}$ & ${-1.57\pm0.89}$ & ${-0.72\pm0.18}$ & 1.41e-01 \\
2017-03-29 & 57841 & ${-0.44\pm0.21}$ &                  &                  & ${-0.68\pm0.18}$ & 1.14e-01 \\
2017-04-19 & 57862 & ${-0.47\pm0.21}$ &                  &                  & ${-0.68\pm0.18}$ & 8.42e-02 \\
2017-06-17 & 57921 & ${-0.66\pm0.21}$ & ${-0.48\pm0.32}$ & ${-1.20\pm0.89}$ & ${-0.67\pm0.18}$ & 2.92e-02 \\
2017-09-19 & 58015 & ${-0.61\pm0.21}$ & ${-0.43\pm0.32}$ &                  & ${-0.52\pm0.16}$ & 3.09e-02 \\
2017-10-21 & 58047 &                  &                  & ${-1.67\pm0.89}$ & ${-0.76\pm0.18}$ & 1.10e-01 \\
2017-11-05 & 58062 & ${-0.52\pm0.21}$ & ${-0.47\pm0.32}$ & ${-0.93\pm0.89}$ & ${-0.57\pm0.18}$ & 2.29e-02 \\
2017-11-22 & 58079 & ${-0.46\pm0.21}$ &                  &                  & ${-0.71\pm0.18}$ & 1.20e-01 \\\addlinespace
2018-01-13 & 58131 & ${-0.36\pm0.21}$ & ${-0.60\pm0.32}$ & ${-1.47\pm0.89}$ & ${-0.67\pm0.18}$ & 1.80e-01 \\
2018-02-19 & 58168 & ${-0.41\pm0.21}$ & ${-0.45\pm0.32}$ & ${-1.07\pm0.89}$ & ${-0.55\pm0.18}$ & 7.71e-02 \\
2018-03-26 & 58203 & ${-0.47\pm0.21}$ & ${-0.34\pm0.32}$ & ${-1.10\pm0.89}$ & ${-0.53\pm0.18}$ & 7.10e-02 \\
2018-04-19 & 58227 & ${-0.32\pm0.21}$ & ${-0.23\pm0.32}$ & ${-0.97\pm0.89}$ & ${-0.40\pm0.18}$ & 1.49e-01 \\
2018-05-16 & 58254 &                  &                  &                  & ${-0.27\pm0.16}$ &          \\
2019-01-26 & 58509 & ${-0.56\pm0.21}$ & ${-0.48\pm0.32}$ &                  & ${-0.52\pm0.16}$ & 7.18e-03 \\
2019-02-15 & 58529 & ${-0.54\pm0.21}$ &                  &                  & ${-0.54\pm0.21}$ &          \\
2019-04-20 & 58593 & ${-0.50\pm0.21}$ & ${-0.84\pm0.32}$ & ${-1.10\pm0.89}$ & ${-0.77\pm0.18}$ & 7.20e-02 \\
2019-06-03 & 58637 & ${-0.33\pm0.21}$ & ${-0.83\pm0.32}$ & ${-1.12\pm0.89}$ & ${-0.71\pm0.18}$ & 1.64e-01 \\
2019-06-24 & 58658 & ${-0.53\pm0.21}$ & ${-0.94\pm0.32}$ &                  & ${-0.74\pm0.16}$ & 1.01e-01 \\
2020-01-08 & 58856 & ${-0.64\pm0.21}$ & ${-1.01\pm0.32}$ & ${-1.69\pm0.89}$ & ${-1.00\pm0.18}$ & 1.09e-01 \\
2020-02-13 & 58892 & ${-0.64\pm0.21}$ & ${-0.91\pm0.32}$ &                  & ${-0.78\pm0.16}$ & 3.67e-02 \\
2020-03-13 & 58921 & ${-0.75\pm0.21}$ & ${-1.05\pm0.32}$ &                  & ${-0.90\pm0.16}$ & 3.32e-02 \\
\bottomrule
\end{tabular}
}}
\tabnote{\textbf{Notes.} (1)~Date in yyyy-mm-dd, (2)~Modified Julian Date, (3)~spectral index between 22~GHz and 43~GHz using single power-law, (4)~spectral index between 43~GHz and 86~GHz using single power-law, (5)~spectral index between 86~GHz and 129~GHz using single power-law, (6)~spectral index for all data at frequencies that could be obtained for each epoch using single power-law, (7)~reduced chi-square of the spectral fitting for ${\alpha_{\rm{whole}}}$. The contents of this table were separated with gap according to the period~1-4 in Figure~\ref{fig:lc_cq}. ${^{*}}$ ${\alpha_{\rm{whole}}}$ has different meanings depending on the epoch: it is the result of the actual fit for the whole band, while, in other cases, this number indicates only a straight line fit between two frequencies.}
\end{table*}
%%%%%%%%%%%%%%%%%%%%%%%%%%%%%%%%%%%%%%%%%%%%%%%%%%%%%%%%%%%%%%%%%%%%%

In Figure~\ref{fig:pl}, the spectra of the variable flux density ${S_{\rm{var}}}$ are shown in green. The adjacent frequency spectral indices~(dashed lines) and the single power-law fitting results~(solid lines) are described in the same way as ${S_{\rm total}}$. For the epoch in which the peak frequency~(${\nu_{\rm{peak}}}$, i.e., the turnover frequency) can be obtained, curved power-law fitting was performed using Equation~(\ref{eq:curved_pl}), and the fitting results are indicated by the gray curves in Figure~\ref{fig:pl}. The results of the spectral analysis using the variable flux density ${S_{\rm{var}}}$ are summarized in Table~\ref{tab:spectral_indices_var}, which lists the spectral indices as in Table~\ref{tab:spectral_indices_CLEAN} and additionally lists ${\nu_{\rm{peak}}}$ and the peak flux density~(${S_{\rm{peak}}}$) values at ${\nu_{\rm{peak}}}$ obtained from the curved power-law fitting using Equation~(\ref{eq:curved_pl}). It should be noted that the spectral results of the curved power-law fitting for the epochs MJD~56379, 56420, 56616, 56650, 56738, 56769, and 56901 of period~1 are excluded as efforts to obtain the best fitting parameters failed. The uncertainties of the spectral indices in Table~\ref{tab:spectral_indices_var} and the uncertainties of ${S_{\rm{var}}}$ are calculated using the error propagation method. The ranges of ${\alpha}$ of the adjacent frequencies are summarized in Table~\ref{tab:range_alpha}.

%%% TABLE %%%%%%%%%%%%%%%%%%%%%%%%%%%%%%%%%%%%%%%%%%%%%%%%%%%%%%%%%%%
%appendix
\begin{table*}[p]
\caption{\label{tab:spectral_indices_var} Spectral indices and results of the curved power-law fitting for the variable flux densities.}
\centering
\resizebox{\textwidth}{!}{%
{\small
\begin{tabular}{lcccccccc}
\toprule
 Date & MJD & ${\alpha_{\rm{22-43}}}$ & ${\alpha_{\rm{43-86}}}$ & ${\alpha_{\rm{86-129}}}$ & ${\alpha_{\rm{whole}}^{*}}$ & ${\chi_{\rm{reduced}}}$ & ${\nu_{\rm{peak}}}$ & ${S_{\rm{peak}}}$\\
  &  &  &  &  &  &  & (GHz) & (Jy)\\
(1) & (2) & (3) & (4) & (5) & (6) & (7) & (8) & (9) \\
\midrule
2013-01-16 & 56308 & ${-0.04\pm 0.48}$ &                   &                   & ${-0.04\pm 0.48}$ &          &                   &                    \\
2013-03-28 & 56379 & ${-0.22\pm 0.51}$ & ${-0.37\pm 0.57}$ &                   & ${-0.29\pm 0.31}$ & 6.69e-02 &                   &                    \\
2013-04-11 & 56393 & ${-0.07\pm 0.51}$ &                   &                   & ${-0.07\pm 0.51}$ &          &                   &                    \\
2013-05-08 & 56420 & ${-0.11\pm 0.51}$ & ${-0.21\pm 0.54}$ & ${-0.21\pm 1.27}$ & ${-0.18\pm 0.28}$ & 4.42e-02 &                   &                    \\
2013-09-24 & 56559 & ${-0.57\pm 0.64}$ &                   &                   & ${-0.57\pm 0.64}$ &          &                   &                    \\
2013-11-20 & 56616 & ${ 0.07\pm 0.56}$ & ${-0.08\pm 0.55}$ & ${-1.20\pm 1.38}$ & ${-0.23\pm 0.32}$ & 1.15e+00 &                   &                    \\
2013-12-24 & 56650 & ${-0.15\pm 0.56}$ & ${ 0.03\pm 0.56}$ & ${-1.69\pm 1.48}$ & ${-0.35\pm 0.34}$ & 9.63e-01 &                   &                    \\
2014-01-27 & 56684 & ${ 0.17\pm 0.66}$ &                   &                   & ${ 0.17\pm 0.66}$ &          &                   &                    \\
2014-02-28 & 56716 & ${ 0.19\pm 0.69}$ & ${-0.10\pm 0.62}$ &                   & ${ 0.04\pm 0.38}$ & 2.42e+00 & ${48.32\pm37.93}$ & ${ 0.92\pm 0.25}$ \\
2014-03-22 & 56738 & ${ 0.29\pm 0.75}$ & ${ 0.13\pm 0.60}$ & ${-1.36\pm 1.49}$ & ${-0.09\pm 0.38}$ & 1.97e+00 &                   &                    \\
2014-04-22 & 56769 & ${ 0.24\pm 1.08}$ & ${ 0.40\pm 0.75}$ &                   & ${ 0.32\pm 0.54}$ & 5.68e-02 &                   &                    \\
2014-09-01 & 56901 & ${ 0.32\pm 0.66}$ & ${-0.04\pm 0.57}$ &                   & ${ 0.14\pm 0.36}$ & 1.19e+00 & ${56.79\pm43.76}$ & ${ 1.08\pm 0.42}$ \\
2014-10-29 & 56959 & ${ 0.14\pm 0.97}$ & ${-0.19\pm 0.82}$ &                   & ${-0.03\pm 0.52}$ & 2.80e+00 & ${41.17\pm22.48}$ & ${ 0.59\pm 0.11}$ \\
2014-11-28 & 56989 & ${ 0.85\pm 1.03}$ &                   &                   & ${ 0.85\pm 1.03}$ &          &                   &                    \\ \addlinespace
2014-12-26 & 57017 & ${ 0.68\pm 0.73}$ & ${ 0.04\pm 0.54}$ &                   & ${ 0.35\pm 0.39}$ & 8.32e-01 & ${63.29\pm30.74}$ & ${ 1.20\pm 0.57}$ \\
2015-02-24 & 57077 & ${ 0.35\pm 0.61}$ & ${-0.19\pm 0.55}$ &                   & ${ 0.07\pm 0.35}$ & 2.58e+00 & ${47.76\pm19.45}$ & ${ 1.19\pm 0.32}$ \\
2015-03-26 & 57107 & ${ 0.17\pm 0.55}$ & ${-0.19\pm 0.54}$ & ${-0.81\pm 1.32}$ & ${-0.18\pm 0.30}$ & 1.16e+00 & ${42.48\pm22.81}$ & ${ 1.28\pm 0.26}$ \\
2015-04-30 & 57142 & ${ 0.49\pm 0.50}$ & ${-0.02\pm 0.47}$ &                   & ${ 0.23\pm 0.29}$ & 1.01e+00 & ${59.52\pm34.53}$ & ${ 1.77\pm 0.75}$ \\
2015-09-24 & 57289 & ${ 0.22\pm 0.53}$ &                   &                   & ${ 0.22\pm 0.53}$ &          &                   &                    \\
2015-10-23 & 57318 & ${ 0.26\pm 0.50}$ &                   &                   & ${ 0.26\pm 0.50}$ &          &                   &                    \\
2015-11-30 & 57356 & ${ 0.26\pm 0.54}$ & ${-0.14\pm 0.51}$ & ${-1.36\pm 1.34}$ & ${-0.22\pm 0.31}$ & 1.51e+00 &                   &                    \\
2015-12-28 & 57384 & ${ 0.31\pm 0.66}$ & ${-0.26\pm 0.59}$ & ${-2.04\pm 1.71}$ & ${-0.38\pm 0.40}$ & 1.49e+00 &                   &                    \\ \addlinespace
2016-03-01 & 57448 & ${ 0.40\pm 0.91}$ & ${ 0.01\pm 0.69}$ & ${-1.47\pm 1.73}$ & ${-0.12\pm 0.45}$ & 1.93e+00 &                   &                    \\
2016-04-24 & 57502 & ${ 0.94\pm 0.81}$ & ${-0.04\pm 0.55}$ &                   & ${ 0.44\pm 0.43}$ & 1.23e+00 & ${59.21\pm17.54}$ & ${ 1.23\pm 0.52}$ \\
2016-08-23 & 57623 & ${ 0.46\pm 0.59}$ &                   &                   & ${ 0.46\pm 0.59}$ &          &                   &                    \\
2016-10-18 & 57679 & ${ 0.22\pm 0.56}$ & ${-0.17\pm 0.54}$ & ${-1.09\pm 1.36}$ & ${-0.20\pm 0.31}$ & 1.34e+00 &                   &                    \\
2016-12-28 & 57750 & ${-0.07\pm 0.59}$ & ${-0.22\pm 0.60}$ & ${-1.82\pm 1.67}$ & ${-0.46\pm 0.38}$ & 7.86e-01 &                   &                    \\
2017-03-29 & 57841 & ${ 0.05\pm 0.67}$ &                   &                   & ${-0.34\pm 0.37}$ & 7.95e-01 & ${32.99\pm10.64}$ & ${ 0.93\pm 0.11}$ \\
2017-04-19 & 57862 & ${ 0.00\pm 0.70}$ &                   &                   & ${-0.32\pm 0.38}$ & 6.27e-01 & ${30.91\pm11.90}$ & ${ 0.86\pm 0.10}$ \\
2017-06-17 & 57921 & ${-0.50\pm 0.68}$ & ${ 0.12\pm 0.69}$ & ${-1.18\pm 1.63}$ & ${-0.33\pm 0.37}$ & 4.81e-01 &                   &                    \\
2017-09-19 & 58015 & ${-0.24\pm 1.01}$ & ${ 0.43\pm 0.83}$ &                   & ${ 0.11\pm 0.50}$ & 2.15e+00 &                   &                    \\
2017-10-21 & 58047 &                   &                   & ${-2.31\pm 2.77}$ & ${-0.27\pm 0.71}$ & 2.75e+00 & ${45.99\pm 2.86}$ & ${ 0.81\pm 0.25}$ \\
2017-11-05 & 58062 & ${ 0.23\pm 1.27}$ & ${ 0.38\pm 0.86}$ & ${-0.69\pm 1.74}$ & ${ 0.13\pm 0.53}$ & 1.41e+00 &                   &                    \\
2017-11-22 & 58079 & ${ 0.62\pm 1.51}$ &                   &                   & ${-0.06\pm 0.67}$ & 2.92e+00 & ${52.17\pm13.32}$ & ${ 0.47\pm 0.12}$ \\ \addlinespace
2018-01-13 & 58131 & ${ 0.55\pm 1.00}$ & ${-0.09\pm 0.72}$ & ${-1.68\pm 1.90}$ & ${-0.16\pm 0.50}$ & 1.94e+00 &                   &                    \\
2018-02-19 & 58168 & ${ 0.21\pm 0.76}$ & ${ 0.08\pm 0.63}$ & ${-0.98\pm 1.49}$ & ${-0.07\pm 0.37}$ & 1.92e+00 &                   &                    \\
2018-03-26 & 58203 & ${-0.08\pm 0.62}$ & ${ 0.19\pm 0.58}$ & ${-1.04\pm 1.37}$ & ${-0.13\pm 0.32}$ & 1.56e+00 &                   &                    \\
2018-04-19 & 58227 & ${ 0.31\pm 0.65}$ & ${ 0.29\pm 0.53}$ & ${-0.88\pm 1.26}$ & ${ 0.09\pm 0.32}$ & 1.77e+00 &                   &                    \\
2018-05-16 & 58254 &                   &                   &                   & ${ 0.26\pm 0.33}$ &          &                   &                    \\
2019-01-26 & 58509 & ${-0.37\pm 0.49}$ & ${-0.10\pm 0.54}$ &                   & ${-0.23\pm 0.29}$ & 2.81e-01 &                   &                    \\
2019-02-15 & 58529 & ${-0.28\pm 0.57}$ &                   &                   & ${-0.28\pm 0.57}$ &          &                   &                    \\
2019-04-20 & 58593 & ${-0.10\pm 0.69}$ & ${-0.60\pm 0.75}$ & ${-0.99\pm 1.90}$ & ${-0.49\pm 0.41}$ & 3.28e-01 & ${26.99\pm 8.20}$ & ${ 0.88\pm 0.09}$ \\
2019-06-03 & 58637 & ${ 0.32\pm 0.70}$ & ${-0.62\pm 0.68}$ & ${-1.05\pm 1.75}$ & ${-0.37\pm 0.40}$ & 1.08e+00 & ${38.77\pm 6.95}$ & ${ 0.99\pm 0.17}$ \\
2019-06-24 & 58658 & ${-0.13\pm 0.72}$ & ${-0.80\pm 0.85}$ &                   & ${-0.47\pm 0.47}$ & 5.93e-01 & ${27.04\pm 6.08}$ & ${ 0.83\pm 0.09}$ \\
2020-01-08 & 58856 & ${ 0.16\pm 2.65}$ & ${-0.81\pm 2.51}$ &                   & ${-0.34\pm 1.58}$ & 1.63e+00 & ${34.44\pm 5.33}$ & ${ 0.20\pm 0.03}$ \\
2020-02-13 & 58892 &                   & ${ 0.24\pm 4.52}$ &                   & ${ 0.24\pm 4.52}$ &          &                   &                    \\
2020-03-13${^{\rm**}}$ & 58921 &                   &                   &                   &                   &          &                   &                    \\
\bottomrule
\end{tabular}
}}
\tabnote{\textbf{Notes.} (1)~Observing date in yyyy-mm-dd, (2)~Modified Julian Date, (3)~spectral index between 22~GHz and 43~GHz using the single power-law, (4)~spectral index between 43~GHz and 86~GHz using the single power-law, (5)~spectral index between 86~GHz and 129~GHz using the single power-law, (6)~spectral index for all data at frequencies that could be obtained for each epoch using single power-law, (7)~the reduced chi-square of the spectral fitting for ${\alpha_{\rm{whole}}}$, (8)~peak frequency, and (9)~flux density at the peak frequency. The contents of this table were separated with gaps according to periods~1-4 in Figure~\ref{fig:lc_cq}. ${^{*}}$ ${\alpha_{\rm{whole}}}$ has different meanings depending on the epoch: it is the result of the actual fit for the whole band, while, in other cases, this number indicates only a straight line fit between two frequencies. ${^{\rm**}}$ No power-law method was applied, as there was only a variable flux density at 22~GHz.}
\end{table*}
%%%%%%%%%%%%%%%%%%%%%%%%%%%%%%%%%%%%%%%%%%%%%%%%%%%%%%%%%%%%%%%%%%%%%
%%% FIGURE %%%%%%%%%%%%%%%%%%%%%%%%%%%%%%%%%%%%%%%%%%%%%%%%%%%%%%%%%%
\begin{figure*}[hbt!]
\centering
\includegraphics[width=\textwidth]{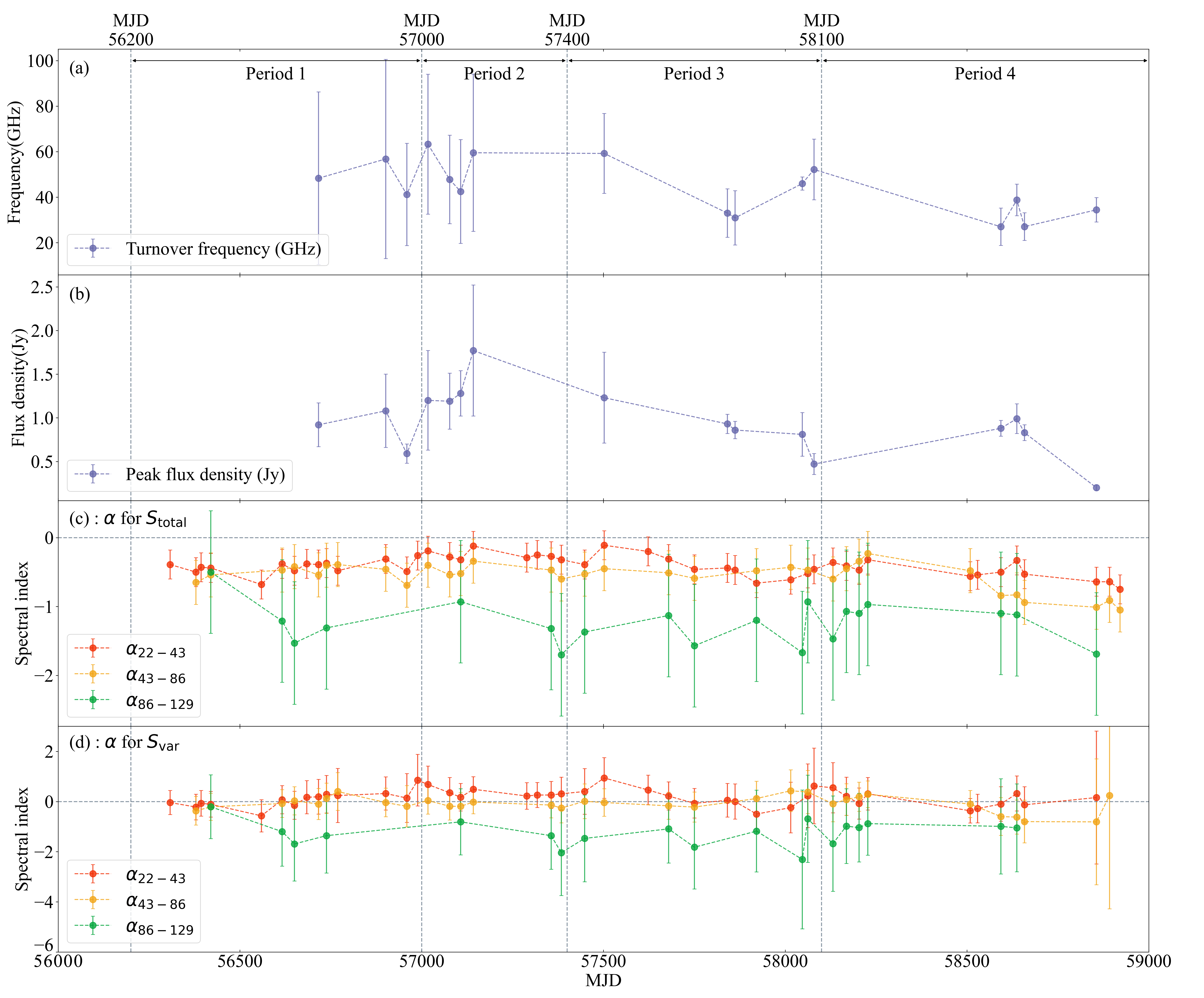}
\caption{(a)~turnover frequencies of the variable flux densities, (b)~peak flux densities at the turnover frequencies of the variable flux densities, (c)~spectral indices of the total CLEANed flux densities, (d)~the spectral indices of the variable flux densities. In~(c) and~(d), the red circles are the spectral indices of 22-43~GHz, the yellow circles are the spectral indices of 43-86~GHz, and the green circles are the spectral indices of 86-129~GHz. Also, the gray horizontal lines denote ${\alpha = 0}$. In~(d), the uncertainty of ${\alpha_{\rm{43-86}}}$ located at the end was too large, and the variability of other data was not properly distinguished, so the maximum value of the y-axis was fixed. \label{fig:peak_freq}}
\end{figure*}
%%%%%%%%%%%%%%%%%%%%%%%%%%%%%%%%%%%%%%%%%%%%%%%%%%%%%%%%%%%%%%%%%%%%%

In Figure~\ref{fig:peak_freq}, the variation of the spectral properties of Table~\ref{tab:spectral_indices_CLEAN} and Table~\ref{tab:spectral_indices_var} are shown as a function of time. In panels (a) and (b), ${\nu_{\rm{peak}}}$ and ${S_{\rm{peak}}}$ are plotted, showing the mild variation of ${\nu_{\rm{peak}}}$ in a range of 27-63~GHz and a rapid increase of the peak flux density from 1.2~Jy to 1.8~Jy at the beginning of period~2, and a gradual decrease to 0.2~Jy afterwards. In panels (c) and (d), the adjacent spectral indices of the ${S_{\rm total}}$ (c) and ${S_{\rm var}}$ (d) are plotted with the spectral indices at 22-43~GHz~(${\alpha_{\rm{22-43}}}$) in red, those at 43-86~GHz~(${\alpha_{\rm{43-86}}}$) are in yellow, those at 86-129~GHz~(${\alpha_{\rm{86-129}}}$) are in green, and the gray horizontal lines for ${\alpha = 0}$. It is clear that the mas-scale total CLEANed flux density ${S_{\rm total}}$ of the source is optically thin in the period and that the variable flux density ${S_{\rm var}}$ is quite flat or inverted at lower frequency bands, indicating that the radio variability of the source is dominated by emission from compact and optically thick regions~(e.g., a radio core).

%%% TABLE %%%%%%%%%%%%%%%%%%%%%%%%%%%%%%%%%%%%%%%%%%%%%%%%%%%%%%%%%%%
%appendix
\begin{table*}[hbt!]
\caption{\label{tab:range_alpha} Ranges of the spectral indices of the adjacent frequency by periods~1-4.}
\centering
\begin{tabular}{l|cccccc}
\toprule
\multicolumn{1}{l|}{} & \multicolumn{6}{c}{${S_{\rm{total}}}$} \\
\cline{2-7}
\multicolumn{1}{l|}{} & \multicolumn{2}{c}{${\alpha_{\rm{22-43}}}$} & \multicolumn{2}{c}{${\alpha_{\rm{43-86}}}$} & \multicolumn{2}{c}{${\alpha_{\rm{86-129}}}$} \\
 & low & high & low & high & low & high \\
\midrule
Period~1 & $-0.68{\pm0.21}$ & $-0.26{\pm0.21}$ & $-0.69{\pm0.32}$ & $-0.39{\pm0.32}$ & $-1.53{\pm0.89}$ & $-0.50{\pm0.89}$ \\
Period~2 & $-0.32{\pm0.21}$ & $-0.12{\pm0.21}$ & $-0.60{\pm0.32}$ & $-0.34{\pm0.32}$ & $-1.70{\pm0.89}$ & $-0.93{\pm0.89}$ \\
Period~3 & $-0.66{\pm0.21}$ & $-0.11{\pm0.21}$ & $-0.59{\pm0.32}$ & $-0.43{\pm0.32}$ & $-1.67{\pm0.89}$ & $-0.93{\pm0.89}$ \\
Period~4 & $-0.75{\pm0.21}$ & $-0.32{\pm0.21}$ & $-1.05{\pm0.32}$ & $-0.23{\pm0.32}$ & $-1.69{\pm0.89}$ & $-0.97{\pm0.89}$ \\
\midrule
\multicolumn{1}{l|}{} & \multicolumn{6}{c}{${S_{\rm{var}}}$} \\
\cline{2-7}
\multicolumn{1}{l|}{} & \multicolumn{2}{c}{${\alpha_{\rm{22-43}}}$} & \multicolumn{2}{c}{${\alpha_{\rm{43-86}}}$} & \multicolumn{2}{c}{${\alpha_{\rm{86-129}}}$} \\
 & low & high & low & high & low & high \\
\midrule
Period~1 & $-0.57{\pm0.64}$ & $0.85{\pm1.03}$ & $-0.37{\pm0.57}$ & $0.40{\pm0.75}$ & $-1.69{\pm1.48}$ & $-0.21{\pm1.27}$\\
Period~2 & $0.17{\pm0.55}$ & $0.68{\pm0.73}$ & $-0.26{\pm0.59}$ & $0.04{\pm0.54}$ & $-2.04{\pm1.71}$ & $-0.81{\pm1.32}$\\
Period~3 & $-0.50{\pm0.68}$ & $0.94{\pm0.81}$ & $-0.22{\pm0.60}$ & $0.43{\pm0.83}$ & $-2.31{\pm2.77}$ & $-0.69{\pm1.74}$\\
Period~4 & $-0.37{\pm0.49}$ & $0.55{\pm1.00}$ & $-0.81{\pm2.51}$ & $0.29{\pm0.53}$ & $-1.68{\pm1.90}$ & $-0.88{\pm1.26}$\\
\bottomrule
\end{tabular}
\end{table*}
%%%%%%%%%%%%%%%%%%%%%%%%%%%%%%%%%%%%%%%%%%%%%%%%%%%%%%%%%%%%%%%%%%%%%

\subsection{Magnetic Field Strength}
In the spectral analysis of the variable flux densities ${S_{\rm{var}}}$ of 4C~+28.07, we found that the synchrotron radiation at low radio frequencies from the source appears to be absorbed by the synchrotron particles, yielding an SSA feature in the spectra with the turnover frequency ${\nu_{\rm{peak}}}$ and the peak flux density ${S_{\rm{peak}}}$. The spectral quantities of the SSA spectrum can be used to constrain the magnetic field strength in the SSA region~(a steady state) following \citet{marscher1983accurate} and \citet{lee2017interferometric} using
\begin{equation}\label{eq:B_SSA}
    B_{\rm{SSA}} = 10^{-5}b(\alpha)S_{\rm{r}}^{-2}\theta_{\rm{r}}^{4}\nu_{\rm{r}}^{5}\left(\frac{\delta}{1+z}\right)^{-1}~({\rm G}),
\end{equation}
where ${\alpha}$ is the spectral index, ${b(\alpha)}$ is the parameter of ${\alpha}$, ${S_{\rm {r}}(=S_{\rm {peak}})}$ is the flux density for the peak frequency, ${\theta_{\rm {r}}}$ is the angular diameter of the emission region, ${\nu_{\rm r}(=\nu_{\rm {peak}})}$ is the peak frequency, ${\delta}$ is the Doppler factor, and ${z}$ is the redshift.

\begin{figure*}[hbt!]
\centering
\includegraphics[width=\textwidth]{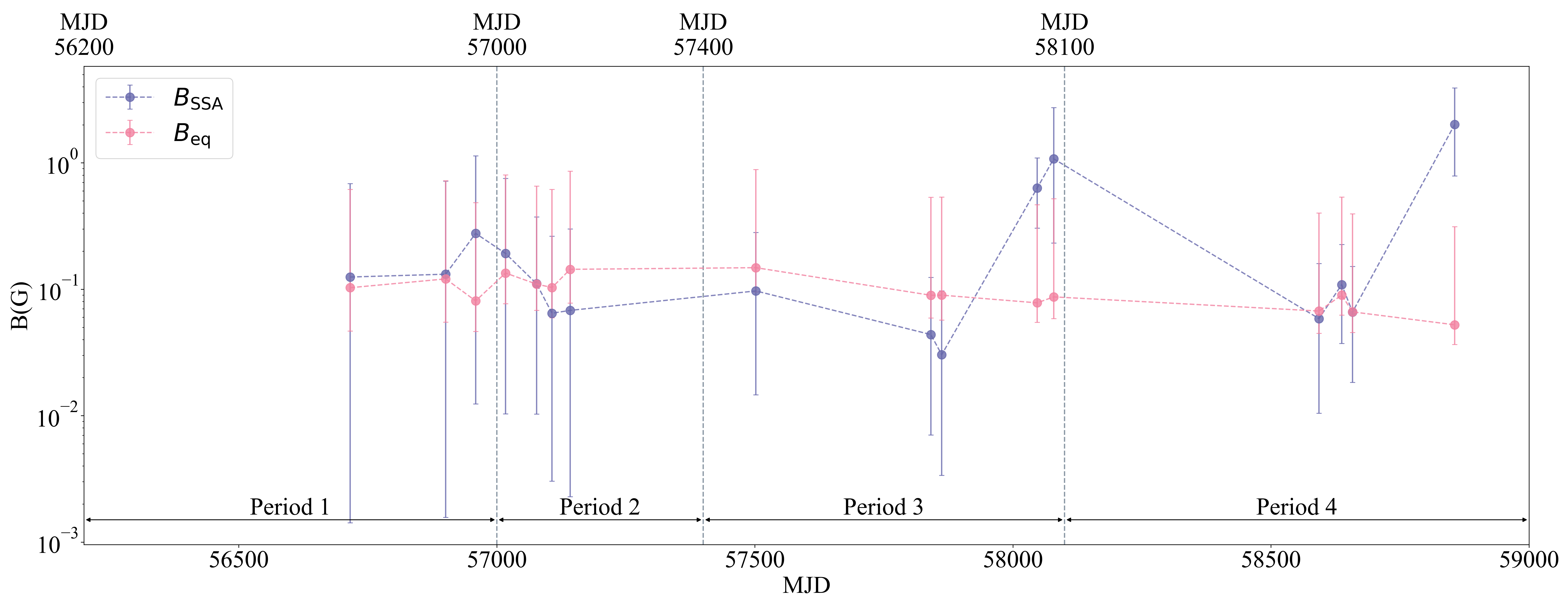}
\caption{The SSA magnetic field strength ${B_{\rm{SSA}}}$ in lavender and the equipartition magnetic field strength ${B_{\rm{eq}}}$ in pink of the SSA region. Each of the values according to the leptonic~${\eta = 1}$ and hadronic~${\eta = 1836}$ scenario was computed as the uncertainty of ${\eta}$ to compute the uncertainty of ${B_{\rm eq}}$.\label{fig:B_SSA}}
\end{figure*}

${b(\alpha)}$ is a factor according to ${\alpha}$ for an optically thin flux density, as described in~\citet{marscher1983accurate}, in this case ${b(\alpha)}=1.8, 3.2, 3.6, 3.8$ for ${\alpha}=0.25, 0.5, 0.75, 1.0$. In order to determine the factor for other values of the optically thin spectral index, we interpolated the factor following \citet{kang2021interferometric}, and we used ${\alpha_{\rm{86-129}}}$ as ${\alpha}$ to compute ${b(\alpha)}$, and in the absence of data at that frequency~(e.g., 86 or 129~GHz), the adjacent spectral index was derived by obtaining the flux density value through Equation~(\ref{eq:curved_pl}). Moreover, we set the lower and upper limit of ${b(\alpha)}$ to 1.8 and 3.8, respectively, to avoid the divergence of ${b(\alpha)}$. The angular diameter ${\theta_{\rm{r}}}$ of the SSA region is obtained using the core sizes at 24~GHz (${\theta_{\rm{FWHM}}}$) from \citet[][see Table 10]{de2023celestial} during the period of 21 July 2015~(MJD~57224) to 22 July 2018~(MJD~58321). The sizes are interpolated in time~(${\theta_{\rm{FWHM,inter}}}$) to determine ${\theta_{\rm{r}}}$ in MJD~57224-58321. For the epochs where the 24~GHz core sizes are not available from \citet{de2023celestial}, we used, $\overline{\theta_{\rm{FWHM}}}$, the mean value of ${\theta_{\rm{FWHM}}}$. The 24~GHz core sizes are further converted, taking into account the jet geometry ${\theta \propto \nu^{-\epsilon}}$~(e.g., $\epsilon = 1$ for a conical jet and $\epsilon=0.5$ for a parabolic jet) following \citet{algaba2017resolving} as
\begin{equation}
    \theta_{\nu_{\rm{peak}}} = \theta_{\rm{24~GHz}} \left( \frac{\nu_{\rm{peak}}}{\nu_{\rm{VLBA}}} \right)^{-\epsilon},
\end{equation}
where $\theta_{\rm{24~GHz}}$ is $\overline{\theta_{\rm{FWHM}}}$ or $\theta_{\rm{FWHM,inter}}$ depending on the epoch, ${\nu_{\rm{VLBA}}}$ is the frequency for each session of the VLBA data~\citep{de2023celestial}, and $\epsilon$ is ${0.66\pm0.07}$~\citep{algaba2017resolving}. And then, we multiply $\theta_{\nu_{\rm{peak}}}$ by 1.8, assuming the emission region is geometrically spherical, and used it as $\theta_{\rm{r}}$~\citep{marscher1977effects}. We used ${\delta} = 23.3$, which is a Doppler factor for 4C~+28.07~\citep{homan2021mojave} and ${z}= 1.206$~\citep{shaw2012spectroscopy}. And we computed the lower uncertainty of $B_{\rm{SSA}}$, following Appendix.~C of~\citet{kang2021interferometric}. The estimated SSA magnetic field strength ${B_{\rm SSA}}$ is summarized in Table~\ref{tab:B_SSA} and shown in Figure~\ref{fig:B_SSA}. We found that the SSA magnetic field strength ranges from 0.030~G to 2.015~G.

Assuming a minimum total energy content in a synchrotron source, an equipartition condition between the magnetic field and particle energy densities is approximated. Taking into account the estimated SSA spectral properties of the source, we are able to estimate the equipartition magnetic field strength ${B_{\rm{eq}}}$
\citep{kataoka2005x, algaba2018exploring, kang2021interferometric} as
\begin{multline}\label{eq:B_eq}
    B_{\rm{eq}} = 123\eta^{\rm{2/7}}(1+z)^{\rm{11/7}}\left( \frac{D_{\rm L}}{100~{\rm{Mpc}}} \right)^{\rm{-2/7}}\left( \frac{\nu_{\rm r}}{5~{\rm{GHz}}} \right)^{\rm{1/7}} \\
    \times\left( \frac{S_{\rm r}}{100~{\rm{mJy}}} \right)^{\rm{2/7}}\left( \frac{\theta_{\rm r}}{0.3~{\rm{''}}} \right)^{\rm{-6/7}}\delta^{\rm{-5/7}}~(\mu {\rm G}),
\end{multline}
where ${\eta}$ is the ratio of energy density between protons and electrons, i.e., if the leptonic scenario then ${\eta = 1}$, or the hadronic scenario then ${\eta=1836}$. Here, we assumed ${\eta\approx100}$, since it is usually an unknown parameter. To preserve the possibility of leptonic or hadronic scenarios, we computed 1 and 1836 as the uncertainty of $\eta$. Moreover, we used ${D_{\rm L} = 8.38}$~Gpc, which is the luminosity distance~\citep{zargaryan2022multiwavelength}. We found that under the assumption of the equipartition condition in the source, ${B_{\rm{eq}}}$ of the SSA region should range from 0.052~G to 0.148~G as listed in Table~\ref{tab:B_SSA}.

%%% TABLE %%%%%%%%%%%%%%%%%%%%%%%%%%%%%%%%%%%%%%%%%%%%%%%%%%%%%%%%%%%
%appendix
\begin{table*}[hbt!]
\caption{\label{tab:B_SSA} Estimated B-field strengths and parameters of SSA regions with the variable fluxes.}
\centering
\begin{tabular}{lccccccccc}
\toprule
Date & MJD & ${b(\alpha)}$ & ${S_{\rm r}}$ & $\theta^{\rm a}_{\rm{r}}$ & ${\nu_{\rm{r}}}$ & ${\delta^{\rm b}}$ & ${z^{\rm c}}$ & ${B_{\rm{SSA}}}$ & ${B_{\rm{eq}}}$\\
 &  &  & (Jy) & (mas) & (GHz) &  &  & (G) & (G)\\
\midrule
2014-02-28 & 56716 & 2.242${^{\rm d}}$ & 0.92${\pm0.25}$ & 0.065${\pm0.034}$ & 48.32${\pm37.93}$ & 23.3 & 1.206 & 0.125${^{+0.560}_{-0.124}}$ & 0.103${^{+0.513}_{-0.056}}$ \\
2014-09-01 & 56901 & 2.222${^{\rm d}}$ & 1.08${\pm0.42}$ & 0.059${\pm0.03}$ & 56.79${\pm43.76}$ & 23.3 & 1.206 & 0.132${^{+0.583}_{-0.130}}$ & 0.121${^{+0.602}_{-0.066}}$ \\
2014-10-29 & 56959 & 2.980${^{\rm d}}$ & 0.59${\pm0.11}$ & 0.072${\pm0.026}$ & 41.17${\pm22.48}$ & 23.3 & 1.206 & 0.277${^{+0.860}_{-0.264}}$ & 0.081${^{+0.402}_{-0.035}}$ \\
2014-12-26 & 57017 & 3.098${^{\rm d}}$ & 1.20${\pm0.57}$ & 0.055${\pm0.018}$ & 63.29${\pm30.74}$ & 23.3 & 1.206 & 0.192${^{+0.561}_{-0.182}}$ & 0.134${^{+0.668}_{-0.058}}$ \\
2015-02-24 & 57077 & 3.414${^{\rm d}}$ & 1.19${\pm0.32}$ & 0.066${\pm0.018}$ & 47.76${\pm19.45}$ & 23.3 & 1.206 & 0.111${^{+0.263}_{-0.100}}$ & 0.110${^{+0.545}_{-0.042}}$ \\
2015-03-26 & 57107 & 3.042 & 1.28${\pm0.26}$ & 0.071${\pm0.025}$ & 42.48${\pm22.81}$ & 23.3 & 1.206 & 0.065${^{+0.198}_{-0.062}}$ & 0.103${^{+0.513}_{-0.044}}$ \\
2015-04-30 & 57142 & 2.763${^{\rm d}}$ & 1.77${\pm0.75}$ & 0.057${\pm0.022}$ & 59.52${\pm34.53}$ & 23.3 & 1.206 & 0.068${^{+0.231}_{-0.066}}$ & 0.144${^{+0.715}_{-0.066}}$ \\
2016-04-24 & 57502 & 3.663${^{\rm d}}$ & 1.23${\pm0.52}$ & 0.049${\pm0.01}$ & 59.21${\pm17.54}$ & 23.3 & 1.206 & 0.097${^{+0.184}_{-0.082}}$ & 0.148${^{+0.736}_{-0.053}}$ \\
2017-03-29 & 57841 & 3.613${^{\rm d}}$ & 0.93${\pm0.11}$ & 0.072${\pm0.015}$ & 32.99${\pm10.64}$ & 23.3 & 1.206 & 0.044${^{+0.080}_{-0.037}}$ & 0.090${^{+0.445}_{-0.030}}$ \\
2017-04-19 & 57862 & 3.475${^{\rm d}}$ & 0.86${\pm0.10}$ & 0.069${\pm0.018}$ & 30.91${\pm11.90}$ & 23.3 & 1.206 & 0.030${^{+0.067}_{-0.027}}$ & 0.090${^{+0.446}_{-0.033}}$ \\
2017-10-21 & 58047 & 3.8${^{\rm e}}$ & 0.81${\pm0.25}$ & 0.085${\pm0.005}$ & 45.99${\pm 2.86}$ & 23.3 & 1.206 & 0.632${^{+0.461}_{-0.327}}$ & 0.078${^{+0.387}_{-0.023}}$ \\
2017-11-22 & 58079 & 3.661${^{\rm d}}$ & 0.47${\pm0.12}$ & 0.064${\pm0.011}$ & 52.17${\pm13.32}$ & 23.3 & 1.206 & 1.076${^{+1.653}_{-0.844}}$ & 0.087${^{+0.432}_{-0.029}}$ \\
2019-04-20 & 58593 & 3.8${^{\rm e}}$ & 0.88${\pm0.09}$ & 0.096${\pm0.019}$ & 26.99${\pm 8.20}$ & 23.3 & 1.206 & 0.059${^{+0.101}_{-0.048}}$ & 0.067${^{+0.334}_{-0.022}}$ \\
2019-06-03 & 58637 & 3.8${^{\rm e}}$ & 0.99${\pm0.17}$ & 0.075${\pm0.009}$ & 38.77${\pm 6.95}$ & 23.3 & 1.206 & 0.109${^{+0.117}_{-0.072}}$ & 0.090${^{+0.446}_{-0.027}}$ \\
2019-06-24 & 58658 & 3.8${^{\rm d, e}}$ & 0.83${\pm0.09}$ & 0.096${\pm0.014}$ & 27.04${\pm 6.08}$ & 23.3 & 1.206 & 0.066${^{+0.085}_{-0.048}}$ & 0.066${^{+0.328}_{-0.021}}$ \\
2020-01-08 & 58856 & 3.8${^{\rm d, e}}$ & 0.20${\pm0.03}$ & 0.081${\pm0.009}$ & 34.44${\pm 5.33}$ & 23.3 & 1.206 & 2.015${^{+1.895}_{-1.228}}$ & 0.052${^{+0.260}_{-0.016}}$ \\
\bottomrule
\end{tabular}
\tabnote{\textbf{Notes.} $^{\rm a}$ referred to~\citet{de2023celestial}, and are obtained by computing the linear interpolation and the proportional expression.
\\$^{\rm b}$ referred to~\citet{homan2021mojave}.
\\$^{\rm c}$ referred to~\citet{shaw2012spectroscopy}.
\\$^{\rm d}$ ${\alpha_{\rm{86-129}}}$ was used as ${\alpha}$ to compute ${b(\alpha)}$, and in the absence of data at that frequency~(e.g., 86 or 129~GHz), the adjacent spectral index was derived by obtaining the flux density value through Equation~(\ref{eq:curved_pl}).
\\$^{\rm e}$ upper limit of ${b(\alpha)}$.
}
\end{table*}
%%%%%%%%%%%%%%%%%%%%%%%%%%%%%%%%%%%%%%%%%%%%%%%%%%%%%%%%%%%%%%%%%%%%%
%%% FIGURE %%%%%%%%%%%%%%%%%%%%%%%%%%%%%%%%%%%%%%%%%%%%%%%%%%%%%%%%%%
%%%%%%%%%%%%%%%%%%%%%%%%%%%%%%%%%%%%%%%%%%%%%%%%%%%%%%%%%%%%%%%%%%%%%

\section{Discussion}\label{sec:5}

\subsection{Multifrequency Variability Characteristics}\label{sec:multifrequency variability characteristics}
We plotted the iMOGABA data~(22-129~GHz) and the ${\gamma}$-ray data, as shown in Figure~\ref{fig:lc}; the data were divided into four periods according to the cadence of the K~band~(22~GHz). Moreover, MJD~55500-56200 was referred to as period~0 because a strong ${\gamma}$-ray flare was detected, but there was no iMOGABA data. First, in period~1, 22 and 43~GHz show a slight ${S_{\rm{total}}}$ rise in the presence of a weak ${\gamma}$-ray flare, while 86 and 129~GHz show no ${S_{\rm{total}}}$ variations due to the small amount of data. Secondly, in period~2, 22, 43, and 86~GHz show relatively gentle hills. The result at 129~GHz is difficult to judge due to the small amount of data, as in period~1. However, the dashed lines from period~1 and period~2 have a very small smooth hill to the flow of 22-86~GHz. Similarly, a weak ${\gamma}$-ray flare is seen in period~2. Thirdly, the 129~GHz data of the period~3 show better variability because the data have a shorter time interval compared to those of the other periods. What stands out in period~3 is that one strong ${\gamma}$-ray flare was detected and that the hill shapes of 22~GHz and 43~GHz are different. At 22~GHz, the distribution of ${S_{\rm{total}}}$ shows a gentle hill shape, but at 43~GHz, the first half of period~3 is skewed. In addition, the result for 86~GHz shows a ${S_{\rm{total}}}$ variability form, similar to that at 43~GHz, whereas the result for 129~GHz has a form similar to that of 22~GHz. What is notable about period~3 is that the peak at 43-86~GHz appeared earlier than the time of the strong ${\gamma}$-ray flares. Therefore, we plan to study in depth a correlation between $\gamma$-ray and radio light curves qualitatively in future work. Finally, all data at 22-129~GHz in period~4 show high peaks, seemingly similar to the period of the strong ${\gamma}$-ray flares.
 
\citet{das2021multi} investigated the correlation between the ${\gamma}$-ray data and the 15~GHz~(OVRO) data for the strong ${\gamma}$-ray flares in 0, 3, and 4, presenting a time lag of about 70~days for the correlation analysis results of the strong ${\gamma}$-ray flare in period~0, and about 135~days for period~4. However, they also found that there was no significant correlation with the 15~GHz data for the strong ${\gamma}$-ray flare of period~3. Therefore, we performed the spectral analysis and computed the magnetic field strength to determine its involvement in this large time lag.

First, we computed ${F_{\rm{var}}}$ to observe how the variability of ${S_{\rm{total}}}$ is represented by the frequency. When we found a slight increase in ${F_{\rm{var}}}$ from 22~GHz to 129~GHz, we could assume that 4C~+28.07 has an optically thin spectrum in period~1-4. Unlike the result of~\citet{lee2020interferometric}, ${F_{\rm{var}}}$ of 4C~+28.07 increased in proportion to the frequency. However, in~\citet{lee2020interferometric}, ${F_{\rm{var}}}$ was obtained by gathering the OVRO data, while here we only investigated the iMOGABA data. Hence, ${F_{\rm{var}}}$ appears to be necessary to analyze with additional frequency data. Secondly, however, in order to examine the peak that appeared before the strong ${\gamma}$-ray flare in period~3, we assumed that the minimum flux emitted by the core of 4C~+28.07 in period~1-4 is ${S_{\rm{q}}}$, with ${S_{\rm{q}}}$ subtracted from all ${S_{\rm{total}}}$ for each frequency and ${S_{\rm{var}}}$ then obtained.

\subsection{Spectral Characteristics}

We obtained ${\alpha}$ of ${S_{\rm{total}}}$ and ${S_{\rm{var}}}$, as detailed in Figure~\ref{fig:peak_freq} and Table~\ref{tab:spectral_indices_CLEAN},~\ref{tab:spectral_indices_var}. Table~\ref{tab:range_alpha} shows that all ${\alpha_{\rm{22-43}}}$, ${\alpha_{\rm{43-86}}}$, and ${\alpha_{\rm{86-129}}}$ values of ${S_{\rm{total}}}$ show ranges lower than 0. However, for ${S_{\rm{var}}}$, there are epochs in which ${\alpha_{\rm{22-43}}}$ and ${\alpha_{\rm{43-86}}}$ are larger than 0. Consequently, in Figure~\ref{fig:peak_freq} (c), all ${\alpha}$ values are below zero. This indicates that the core of 4C~+28.07 played a dominant role in the iMOGABA observation period. This is the reason why we obtained ${S_{\rm{var}}}$ to examine parts other than the core. In Figure~\ref{fig:peak_freq} (d), the values of ${\alpha_{\rm{22-43}}}$ and ${\alpha_{\rm{43-86}}}$ are greater than or equal to zero. This shows that there have been radio emissions in the variable regions, causing an effect. Furthermore, this is more pronounced in the absence of strong ${\gamma}$-ray flares, such as in periods~2-3, and given the presentations in~\citet{das2021multi} and~\citet{zargaryan2022multiwavelength}, 4C~+28.07 can be considered with a two-zone model.

Subsequently, we obtained ${\nu_{\rm{peak}}}$ and ${S_{\rm{peak}}}$ for ${S_{\rm{var}}}$, as presented in Figure~\ref{fig:peak_freq} and Table~\ref{tab:spectral_indices_var}. However, we could not obtain a large number of ${S_{\rm{peak}}}$ values by period. Nevertheless, what is noticeable is that periods~2 and 3 show high ${\nu_{\rm{peak}}}$ values with high ${S_{\rm{peak}}}$ outcomes in the absence of a strong ${\gamma}$-ray flare. This trend appears to require a correlation analysis between the radio data of various frequencies. Because only iMOGABA data and ${\gamma}$-ray data were used in this paper, we decided to determine the distribution of the magnetic field strength levels in the SSA region that appears through ${\nu_{\rm{peak}}}$ in the iMOGABA data.

\subsection{Magnetic Characteristics}
We computed ${B_{\rm{SSA}}}$ through Equation~(\ref{eq:B_SSA}), with the results presented in Table~\ref{tab:B_SSA}. The uncertainties of ${B_{\rm{SSA}}}$ are large, likely due to the large uncertainties in ${\theta_{\rm r}}$ and ${\nu_{\rm{r}}}$. In Figure~\ref{fig:B_SSA}, ${B_{\rm{eq}}}$ is largely consistent with ${B_{\rm{SSA}}}$ within the uncertainty for almost all data points. Considering that ${B_{\rm{SSA}}}$ is mostly measured in the ${\gamma}$-ray quiescent periods, the SSA region in the source is not significantly deviated from the equipartition condition in the quiescent period. In addition, the average values of ${B_{\rm{eq}}}$ and ${B_{\rm{SSA}}}$ are 0.098~G and 0.319~G, respectively, which are similar results to ${B_{\rm{BlobII}}=0.1~{\rm G}}$, the magnetic field strength of the Blob II in a quiescent period for the two-zone model, from~\citet{das2021multi}.

%%% FIGURE %%%%%%%%%%%%%%%%%%%%%%%%%%%%%%%%%%%%%%%%%%%%%%%%%%%%%%%%%%
\begin{figure*}[hbt!]
\centering
\includegraphics[width=\textwidth]{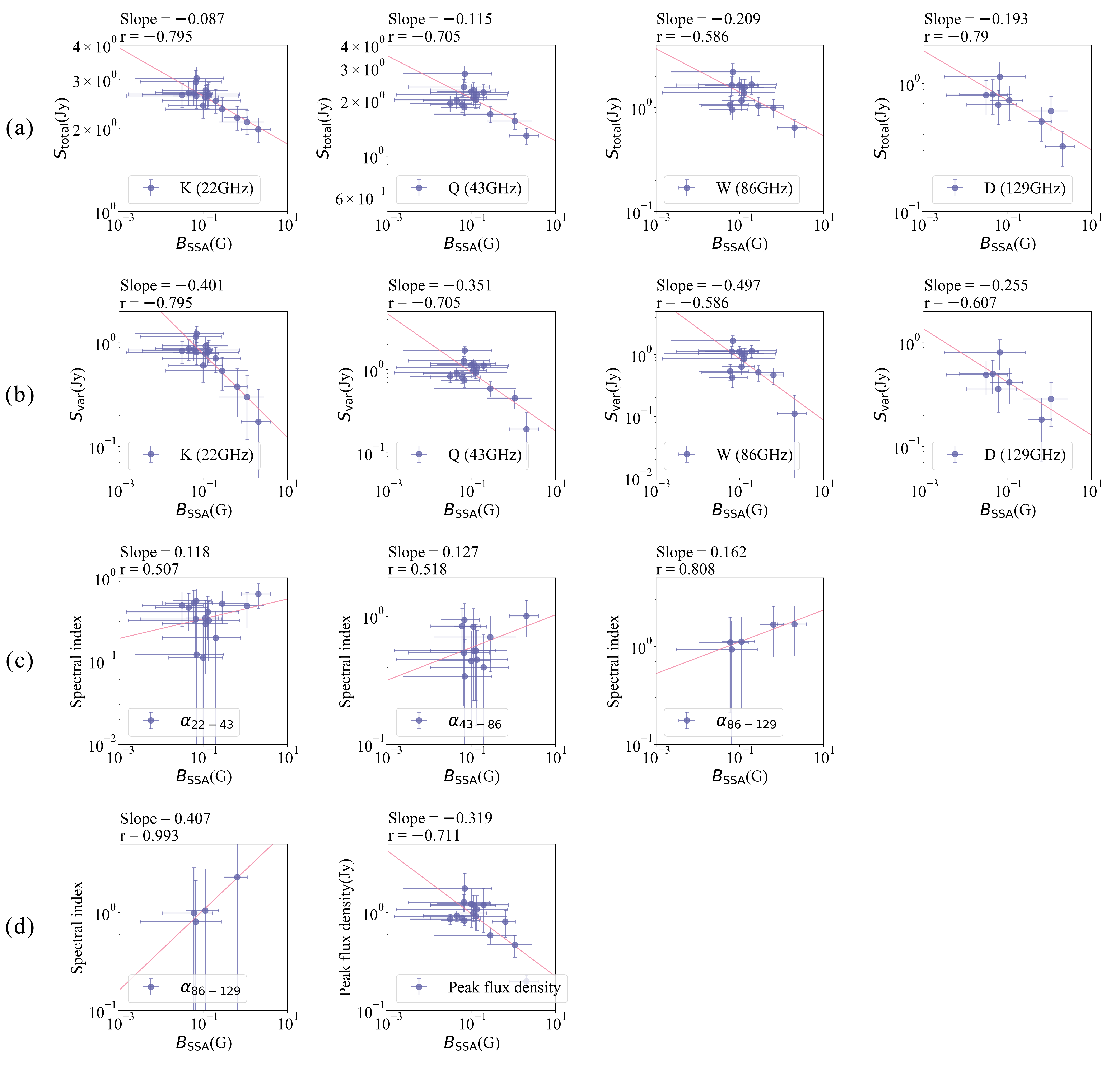}
\caption{Plotted in log-log scale. Lavender color represent the values of each parameter, and pink are line through the least square method between data points. The above `slope' of each graph is the result of the least square method, and `r' is the result of the Pearson correlation coefficient.
(a)~Comparison between ${B_{\rm{SSA}}}$, and ${S_{\rm{total}}}$ of 22, 43, 86, and 129~GHz in order from the left.
(b)~Comparison between ${B_{\rm{SSA}}}$, and ${S_{\rm{var}}}$ of 22, 43, 86, and 129~GHz in order from the left.
(c)~Comparison between ${B_{\rm{SSA}}}$, and the absolute values, because all the values are negative, of ${\alpha_{\rm{22-43}}}$, ${\alpha_{\rm{43-86}}}$, and ${\alpha_{\rm{86-129}}}$ for ${S_{\rm{total}}}$ in order from the left. 
(d)~The left plot is comparison between ${B_{\rm{SSA}}}$ and ${\alpha_{\rm{86-129}}}$ for ${S_{\rm{var}}}$. And the right is comparison between ${B_{\rm{SSA}}}$ and ${S_{\rm{peak}}}$.
\label{fig:B_with_others}}
\end{figure*}
%%%%%%%%%%%%%%%%%%%%%%%%%%%%%%%%%%%%%%%%%%%%%%%%%%%%%%%%%%%%%%%%%%%%%

Furthermore, we compared the relationship between ${B_{\rm{SSA}}}$ and other parameters in Figure~\ref{fig:B_with_others}. In Figure~~\ref{fig:B_with_others}, we investigated the relationship between ${B_{\rm{SSA}}}$ and other parameters such as the peak flux density $S_{\rm peak}$, the spectral indices of $\alpha_{22-43}$, $\alpha_{43-86}$, and $\alpha_{86-129}$, and ${S_{\rm{total}}}$ and ${S_{\rm{var}}}$ at 22-129~GHz. In order to investigate the relationships and the correlations between each parameter, the slopes of the pink lines on the log-log graph were computed using the least square method, and the Pearson correlation coefficient results~(${r}$) are shown above each graph. ${S_{\rm{peak}}}$, ${S_{\rm{total}}}$, and ${S_{\rm{var}}}$ show the inversely proportional correlations. The absolute values of ${\alpha}$ for the adjacent frequencies also present proportional relationships with ${B_{\rm{SSA}}}$.

\subsection{Quiescent and Variable Emission Regions}

Our spectral model of the quiescent and variable contributions may imply two independent components of emission regions in mas-scales probed by the KVN observations. This leads to a possibility of different physical properties for those components, e.g., $B$, $\theta$, etc. Our clear evidence for the components is their spectral properties, i.e., the optically thin spectrum for the quiescent component, and optically thick or SSA spectra for the variable component, indicating that the variable emission region is most likely compact like a core while the quiescent emission region may not be, compared with the variable emission region, leading to a conclusion that a core size from the KVN observations may not be representing the size of the compact variable component, and hence a higher spatial resolution observation is required for constraining the size. Therefore we used the core sizes from the 24~GHz VLBA observations~\citep{de2023celestial} in order to investigate the true size of the variable SSA regions.

Our spectral model of the quiescent and variable contributions may also lead to two magnetic field components. However, we are not sure if the magnetic field strength of the quiescent component is far different from $B_{\rm SSA}$ for the variable compact region or not. And it may be noted that we are not able to constrain the B-field strength of the quiescent component using the SSA analysis. This may lead to future works with high spatial resolution observations with a wide range of frequencies for studying spectral properties~(hence B-field, etc.) of multiple components in the compact emission regions of the relativistic jets.

Moreover, the interpretation of ${S_{\rm var}}$ varies depending on what criteria we set for the quiescent flux or epoch. If the observations of the iMOGABA data had ended earlier, we may obtain an overestimated ${S_{\rm q}}$. For a quantitative investigation, we assumed three selected quiescent spectra, obtained from the local minima on MJD 56738, 57448, and 58062, to be the quiescent spectrum of the source. The selected quiescent flux densities are all brighter than the current ones, as listed in Table~\ref{tab:assuming_qui}, i.e., the ratio $r_{S_{\rm q, band}}$ of the selected local minima flux density to that of the current quiescent flux density is larger than the unity. We estimated a ratio $r_{\alpha_{\rm q}}$ of the spectral index of the selected quiescent spectrum to that of the current quiescent spectrum, as listed in Table~\ref{tab:assuming_qui}, indicating they are flatter than the current spectral index of $-0.973$ but still optically thin~(${\alpha_{\rm{q}}}$ are $-0.558$, $-0.634$, and $-0.572$ at MJD~56738, 57448, and 58062, respectively). In order to investigate the effects on the results, i.e., ${\nu_{\rm{peak}}}$, ${S_{\rm{peak}}}$, ${B_{\rm{SSA}}}$, and ${B_{\rm{eq}}}$, we estimated the ratio of them, again as listed in Table~\ref{tab:assuming_qui}. We found that ${\nu_{\rm{peak}}}$ and ${S_{\rm{peak}}}$ become smaller by 10-60\%. However, the effects result in increasing ${B_{\rm{SSA}}}$ up to about 5~times higher. Therefore, it is possible that the SSA magnetic field strength would have been more enhanced, depending on the multiwavelength light curves available.

%%% TABLE %%%%%%%%%%%%%%%%%%%%%%%%%%%%%%%%%%%%%%%%%%%%%%%%%%%%%%%%%%%
\begin{table}[hbt!]
\caption{\label{tab:assuming_qui} Investigation of the ratio of parameters that vary according to the determination criteria of ${S_{\rm q}}$.}
\centering
\begin{tabular}{ccccc}
\toprule
                           & MJD    & MJD    & MJD     \\
                           & 56738  & 57448  & 58062   \\
\midrule
%$r_{\alpha_{\rm{22-43}}}$  &  1.79  & -12.52 & -0.20   \\
%$r_{\alpha_{\rm{43-86}}}$  &  7.55  &  3.41  & -0.67   \\
%$r_{\alpha_{\rm{86-129}}}$ & -2.24  & -0.10  &  1.62   \\
$r_{S_{\rm{q, 22GHz}}}$    &  1.40  &  1.31  &  1.21   \\
$r_{S_{\rm{q, 43GHz}}}$    &  1.80  &  1.66  &  1.41   \\
$r_{S_{\rm{q, 86GHz}}}$    &  2.81  &  2.37  &  2.10   \\
$r_{S_{\rm{q, 129GHz}}}$   &  2.71  &  2.24  &  2.37   \\
$r_{\alpha_{\rm{q}}}$      &  0.57  &  0.65  &  0.59   \\
%\midrule
$r_{\nu_{\rm{peak}}}$      &  0.75  &  0.91  &  0.82   \\
$r_{S_{\rm{peak}}}$        &  0.44  &  0.69  &  0.66   \\
$r_{B_{\rm{SSA}}}$         &  4.85  &  4.25  &  3.83   \\
$r_{B_{\rm{eq}}}$          &  0.61  &  0.80  &  0.72   \\
\bottomrule
\end{tabular}
\tabnote{\textbf{Notes.} All values are the average values of the ratio calculated with a denominator assuming the minimum value of each frequency as ${S_{\rm q, min}}$, and a numerator ${S_{\rm q, new}}$ assuming each new criterion~(i.e., $\frac{S_{\rm q, new}}{S_{\rm q, min}}$).}
\end{table}
%%%%%%%%%%%%%%%%%%%%%%%%%%%%%%%%%%%%%%%%%%%%%%%%%%%%%%%%%%%%%%%%%%%%%%

\subsection{Comparison with Other Sources}

We found quite interesting to see that the magnetic field $B_{\rm SSA}$ calculated seems to be quite in agreement with the equipartition field $B_{\rm eq}$ within the uncertainties, while other sources studied with the iMOGABA~(e.g, 1633+382 in~\citet{algaba2018exploring}; OJ~287 in~\citet{lee2020interferometric}; 4C~+29.45 in~\citet{kang2021interferometric}, etc., generally show a significant deviation from equipartition). In fact, we have found that the ratio of $B_{\rm eq}/B_{\rm SSA}$ for the other sources is larger than the unity, i.e., up to $B_{\rm eq}/B_{\rm SSA}\le 10^4$ for 1633+382 and OJ~287, and up to $B_{\rm eq}/B_{\rm SSA}\le 10^2$ for 4C~+29.45, whereas for 4C~+28.07 it is $B_{\rm eq}/B_{\rm SSA}\approx 1$ within the uncertainty. In order to understand what is the main parameters for determining the ratio, we investigated the ratio with Equations~(\ref{eq:B_SSA}) and~(\ref{eq:B_eq}) and found that the ratio mainly depends on $\nu_{\rm r}$, $\theta_{\rm r}$, and $S_{\rm r}$ as $B_{\rm eq}/B_{\rm SSA}\propto \nu_{\rm r}^{-34/7}~\theta_{\rm r}^{-34/7}~S_{\rm r}^{16/7}$. This implies that the SSA region in 4C~+28.07 has relatively higher turnover frequencies~($\nu_{\rm r}$), is larger in angular size~($\theta_{\rm r}$), and fainter~($S_{\rm r}$) at the turnover frequency, compared with those in other sources.

\section{Summary}\label{sec:6}

We present the results of the aforementioned analysis of the 22, 43, 86, and 129~GHz data of blazar 4C~+28.07 from the iMOGABA program using the KVN. The period of the iMOGABA data ranged from 16 January 2013 to 13 March 2020~(MJD~56308-58921), and four strong ${\gamma}$-ray flares were observed by the Fermi-LAT within that period. Therefore, we divided the period into four sections, according to the cadence of the 22~GHz data. In addition, although a strong ${\gamma}$-ray flare was observed, the period before the iMOGABA program progressed is referred to as period~0. Furthermore, we obtained ${S_{\rm{var}}}$ to investigate the characteristics of variable emission regions, and the spectral indices and the magnetic field strengths for ${S_{\rm{var}}}$ were computed to identify the characteristics of each period. The characteristics of each period divided in this way were as follows.
\begin{itemize}
    \item Period~0~(MJD~55000-56200): A strong ${\gamma}$-ray flare was observed but is not applicable to the entire observation period of the iMOGABA program.
    \item Period~1~(MJD~56200-57000): This period gradually declined with a high ${S_{\rm{total}}}$ value, estimated to be due to a strong ${\gamma}$-ray flare in period~0. A weak ${\gamma}$-ray was also detected. The ${S_{\rm{total}}}$ outcomes show optically thin spectra, and the ${S_{\rm{var}}}$ outcomes have epochs with the presumptive ${\nu_{\rm{peak}}}$. The variability of ${\alpha}$ due to the adjacent frequencies was also more noticeable in ${S_{\rm{var}}}$ than in ${S_{\rm{total}}}$, where ${\alpha_{\rm{22-43}}}$ and ${\alpha_{\rm{43-86}}}$ were close to or higher than 0, while ${\alpha_{\rm{86-129}}}$ remained significantly lower than 0 after the strong ${\gamma}$-ray flare of period~0. ${B_{\rm{SSA}}}$ and ${B_{\rm{eq}}}$ were also computed, considering the leptonic and hadronic scenarios of ${\eta}$, it is shown that ${B_{\rm{SSA}}}$ and ${B_{\rm{eq}}}$ correspond to an equipartition condition.
    \item Period~2~(MJD~57000-57400): During this period, one weak ${\gamma}$-ray flare was detected, as in period~1; however, the data at 22-86~GHz showed ${S_{\rm{total}}}$ variability in the form of a hill. As in period~1, ${\nu_{\rm{peak}}}$ could be obtained with ${S_{\rm{var}}}$. ${\nu_{\rm{peak}}}$ was not noticeable in periods~1-4, but ${S_{\rm{peak}}}$ was computed and found to be the highest compared to other periods. In the first half, ${B_{\rm{eq}}}$ is almost higher than ${B_{\rm{SSA}}}$, considering without the uncertainty. Moreover, throughout this period, ${\alpha_{\rm{22-43}}}$ remained optically thick, and ${B_{\rm{SSA}}}$ gradually declined, albeit with the large measurement uncertainties.
    \item Period~3~(MJD~57400-58100): A strong ${\gamma}$-ray flare was detected later in this period. On the other hand, the ${S_{\rm{total}}}$ variability of 22~GHz showed a gentle hill shape, but 43~GHz showed a peak point earlier than the time of the strong ${\gamma}$-ray flare. Looking at ${\alpha}$ of the adjacent frequencies of ${S_{\rm{var}}}$, we expect the variability of ${\alpha_{\rm{22-43}}}$ and ${\alpha_{\rm{43-86}}}$ to be correlated with the time of the strong ${\gamma}$-ray flare and accordingly estimate the necessity of conducting multifrequency analysis with more radio data. ${B_{\rm{SSA}}}$ became stronger after the strong ${\gamma}$-ray flare appeared.
    \item Period~4~(MJD~58100-59000): A strong ${\gamma}$-ray flare was detected in the first half of this period, and the data at 22-129~GHz in all cases showed similar levels of variability, although there were very few data points during the flare period. Also during this period, ${\alpha_{\rm{22-43}}}$ tended to be lower than 0 when the strong ${\gamma}$-ray flare appeared, as in period~3. ${B_{\rm{SSA}}}$ could not be obtained when the strong ${\gamma}$-ray flare appeared.
\end{itemize}

As a final point, we estimate that it is necessary to analyze 4C~+28.07 with radio data of various frequencies; the characteristics correlated with it will be analyzed later through a multifrequency analysis. Additionally, we expect that with the spectral indices between adjacent frequencies of the ${S_{\rm{var}}}$ obtained from the VLBI observations, the target source will allow us to determine the possibility of a two-zone model involving the ${\gamma}$-ray flare less.

%%% ACKNOWLEDGMENTS (IF ANY) %%%%%%%%%%%%%%%%%%%%%%%%%%%%%%%%%%%%%%%%

\acknowledgments
We are grateful to the staff of the KVN, who helped us to operate the array and to correlate the data. The KVN is a facility operated by KASI (Korea Astronomy and Space Science Institute). The KVN observations and correlations are supported through high-speed network connections among the KVN sites provided by KREONET (Korea Research Environment Open NETwork), which is managed and operated by KISTI (Korea Institute of Science and Technology Information). This work has made use of Fermi-LAT data supplied by~\citet{abdollahi2023fermi}, \url{https://fermi.gsfc.nas a.gov/ssc/data/access/lat/LightCurveRepository}. This work was supported by a grant from the National Research Foundation of Korea (NRF) funded by the Korean government (MIST) (2020R1A2C2009003).

%%% APPENDICES (IF ANY) %%%%%%%%%%%%%%%%%%%%%%%%%%%%%%%%%%%%%%%%%%%%%
\appendix
\section{Imaging}\label{appendix_imaging}
\subsection{Script Using the CLEAN of the DIFMAP}
\begin{figure*}[hbt!]
\centering
\includegraphics[width=\textwidth]{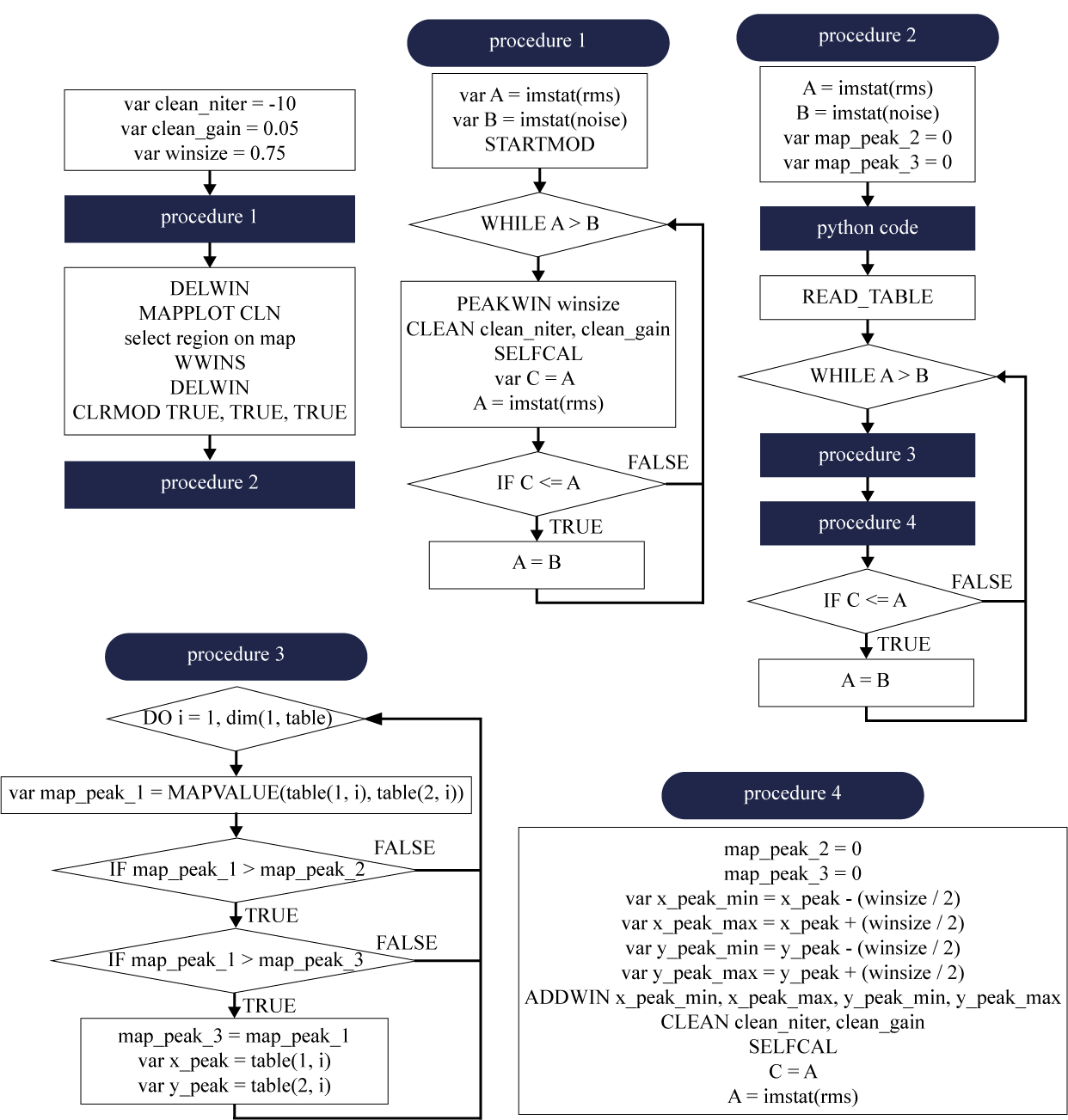}
\caption{Flow chart of the script using the CLEAN of DIFMAP. The round-ended indigo figure is the title of each procedure, and the angled indigo rectangle is the location at which the procedures operate.\label{fig:flowchart_of_script}}
\end{figure*}

In the CLEAN procedure of DIFMAP, the quality of the CLEANed images may depend on the user's proficiency and experience. Therefore, we investigated whether one can proceed with the CLEAN procedure under a more objective standard and perform the CLEAN procedure more quickly and accurately using a script. A main consideration of the script is that CLEAN procedures should be possible regardless of the source morphology (e.g., the orientation and shape of relativistic jets of AGNs) as the script works. Below we describe the script in more detail. A simple flow of the script is as follows with a detailed flowchart shown in Figure~\ref{fig:flowchart_of_script}: 
\begin{enumerate}
    \item CLEANing the whole area of a map
    \item Selecting areas on the residual map
    \item CLEANing the selected areas
    \item Deciding whether or not the CLEAN results are suitable
\end{enumerate}
%%% FIGURE %%%%%%%%%%%%%%%%%%%%%%%%%%%%%%%%%%%%%%%%%%%%%%%%%%%%%%%%%%
%%%%%%%%%%%%%%%%%%%%%%%%%%%%%%%%%%%%%%%%%%%%%%%%%%%%%%%%%%%%%%%%%%%%%
As shown in Figure~\ref{fig:flowchart_of_script}, the main procedures in the script are based on DIFMAP instructions. Initially, the \texttt{niter} and \texttt{gain} variables of CLEAN are set, as is the variable \texttt{winsize}. Here, \texttt{winsize} refers to the size of the selection area that is automatically created when the script runs. We used -10, 0.05, and 0.75 for \texttt{niter}, \texttt{gain}, and \texttt{winsize}, respectively, but for, the 129~GHz data, we used 0.01 for \texttt{gain} because the data had to be CLEANed delicately. Here, 0.75 was selected as the \texttt{winsize} because 1.0 did not produce satisfactory results for CLEAN, and 0.5 was too small. Accordingly, an exceedingly long time was needed to converge the CLEANing. In addition, \texttt{winsize} can be used flexibly by the user depending on the source to be imaged.

After setting up the initial parameters, the script proceeds to procedure~1, i.e., storing the current rms value and noise value of the image as variables A and B, respectively, and executing \texttt{STARTMOD}. We attempted to use \texttt{imstat(noise)} and \texttt{peak(flux, max)} with variable B as parameters to operate the \texttt{while} iteration statement in procedure~1. However, the results of the two parameters were similar. Nevertheless, \texttt{imstat(noise)} produced results slightly faster. Therefore, we decided to use \texttt{imstat(noise)}. Moreover, while the value of A is greater than that of B, a CLEAN window is located in the area including the peaks of the residual map. After performing CLEAN and \texttt{SELFCAL} once, the previous rms value of the residual map is stored as variable C, and the rms value resulting from the CLEAN process is newly stored as variable A. If the newly stored rms value A is greater than or equal to the previous rms value C, the previous noise value B is stored as the rms variable A instead. If the newly stored rms value A is less than the previous rms value C, the creation of the CLEAN window is repeated.

At this point, before moving on to procedure~2, the script deletes all windows created in procedure~1 (\texttt{DELWIN}) and displays the CLEANed map on the screen (\texttt{MAPPLOT CLN}). The user may want to set the CLEAN windows directly in the area they want to analyze. The coordinate values of the selected windows are written in a text file using \texttt{WWINS}. The windows are deleted again with \texttt{DELWIN}, and the CLEANed result from procedure~1 is initialized with \texttt{CLRMOD}. The script then moves on to procedure~2.

Procedure~2 takes a form similar to that of procedure 1. The difference is that it proceeds only within the area selected by the user, and CLEAN is carried out by finding the peak flux density on the residual map. The Python code described in procedure~2 serves to store all coordinate values inside the area selected by the user in the form of a table. The following \texttt{READ\_TABLE} is the command to read this table. 

Procedure 3 is the process of finding the coordinates of the brightest peak by comparing all map coordinates using \texttt{MAPVALUE}. Procedure~4 is the process of performing CLEAN and SELFCAL by making the square window \texttt{ADDWIN} based on the coordinates found. Afterwards, the script is completed through the repetition statement of procedure~2. After the completion of the procedure~2, the results can be confirmed.

\subsection{Verification of the Script}
\begin{figure*}[hbt!]
\centering
\includegraphics[width=\textwidth]{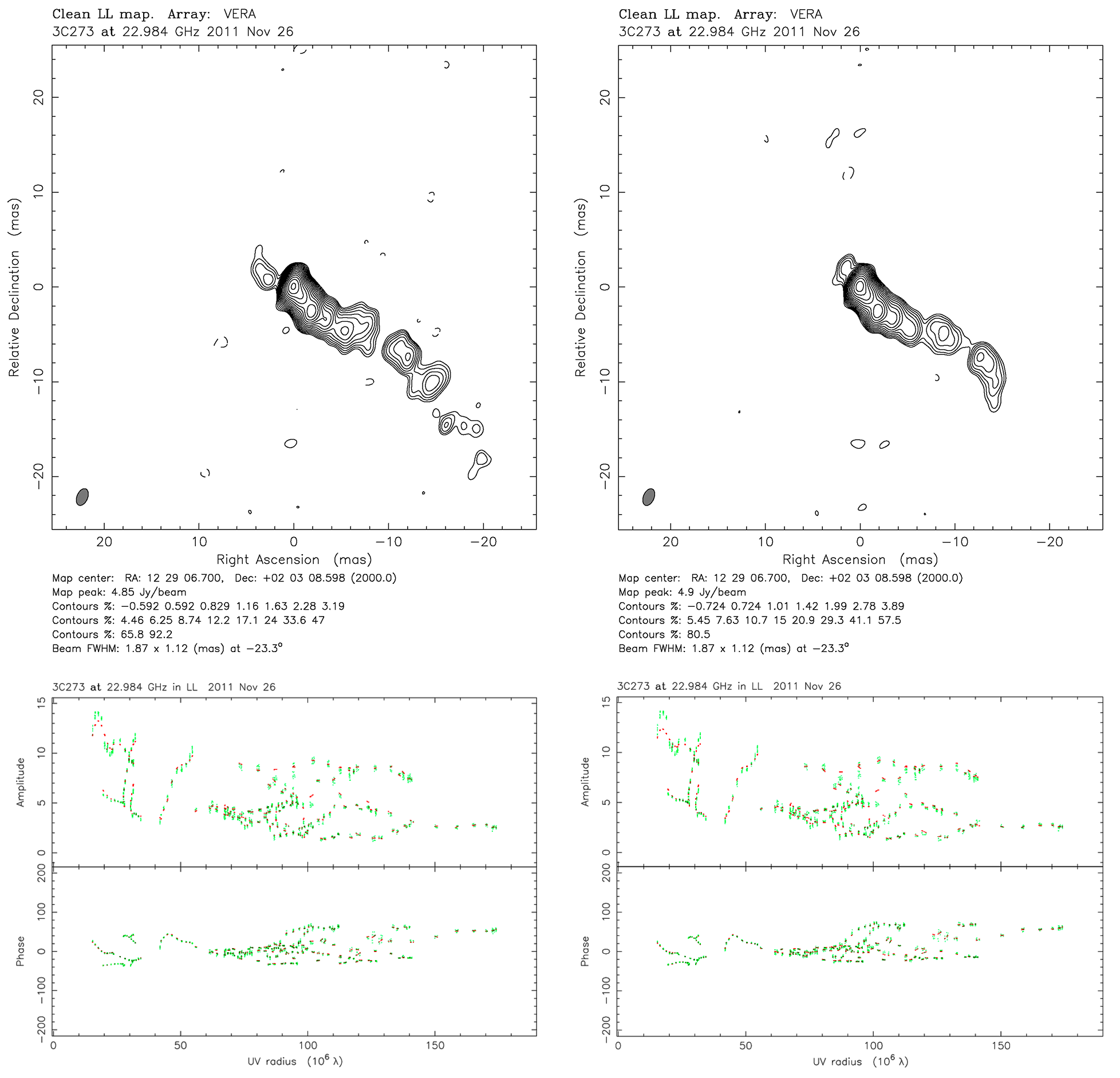}
\caption{CLEANed results of 3C~273 observed by KVN and VERA (KaVA), on 2011 November 26. (Upper)~CLEANed images using the script (left) and being processed manually (right). (Lower)~The visibility amplitude (top) and phase (bottom) of the CLEANed images in the same order as the CLEAN maps, with data in green and CLEAN models in red.\label{fig:clean_result_3c273}}
\end{figure*}

To verify that the script produces proper and accurate results, we used the VLBI data of 3C~273~(with complex jet structures) obtained from the KVN and VLBI Exploration of Radio Astrometry~(VERA) Array~(KaVA). As shown in \citet{niinuma2014vlbi}, 3C~273 has a jet structure that extends to the east-southern direction in the sky. When the CLEAN procedure is performed manually, it is difficult to obtain the faintest structure of the jet properly. Therefore, we expected that the script would be effective if the faintest structure of the jet was properly CLEANed through the script. Before running the script, we set \texttt{niter} and \texttt{gain} to -10 and 0.05, respectively. Consequently, a CLEAN map from the script shows the faintest structure of the jet well. The CLEANed images of 3C~273 are shown in Figure~\ref{fig:clean_result_3c273} for comparison. In the lower left panel (a radplot result of the CLEAN image using the script), the UV data at a UV radius range of about 10~M$\lambda$ to 25~M$\lambda$ correspond to the extended structures of the jet, with the green data representing the actual observation data and with the red data representing the CLEAN models. How consistent the green data and the red data shown in the radplot are is the main criterion for determining how well CLEAN has been performed. It is clear that the CLEANing of the script outperforms manual CLEANing.

%%% FIGURE %%%%%%%%%%%%%%%%%%%%%%%%%%%%%%%%%%%%%%%%%%%%%%%%%%%%%%%%%%
%%%%%%%%%%%%%%%%%%%%%%%%%%%%%%%%%%%%%%%%%%%%%%%%%%%%%%%%%%%%%%%%%%%%%

%% image quality factors
In order to compare the results between scripting and manual CLEANing further, we computed the image quality factor of the CLEANed images of 4C~+28.07, following \citet{lobanov2006dual} and \citet{lee2016interferometric},
\begin{equation}
    {\xi_{\rm r} = s_{\rm r} / s_{\rm r,exp}},
\end{equation}
the ratio of ${|s_{\rm r}|}$~(the maximum absolute flux density in the CLEANed image, using the command \texttt{peak(flux, abs)} in DIFMAP) to ${|s_{\rm r, exp}|}$~(the expectation of ${|s_{\rm r}|}$). ${|s_{\rm r, exp}|}$ can be computed by
\begin{equation}
    {|s_{\rm r, exp}| = \sigma_{\rm r} \left[\ln{\left(\frac{N_{\rm pix}}{\sqrt{2 \pi} \sigma_{\rm r}}\right)}\right]^{\rm 1/2}},
\end{equation}
where ${\sigma_{\rm r}}$ is the image noise rms in the residual image of the source~(using the command \texttt{imstat(rms)} in DIFMAP), and ${N_{\rm pix}}$ is the total number of pixels for reduced data~(i.e., regardless of the residual image or the CLEANed image). If ${\xi_{\rm r} \rightarrow 1}$, the residual noise converges to Gaussian noise. In \citet{lobanov2006dual}, ${\xi_{\rm r}}$ should be close to unity if CLEANing is done properly. If ${\xi_{\rm r} > 1}$, the emission of the source is not fully CLEANed, and if ${\xi_{\rm r} < 1}$ and closer to 0, the imaging has been over-handled.

\begin{figure}[hbt!]
\centering
\includegraphics[width=0.35\textwidth]{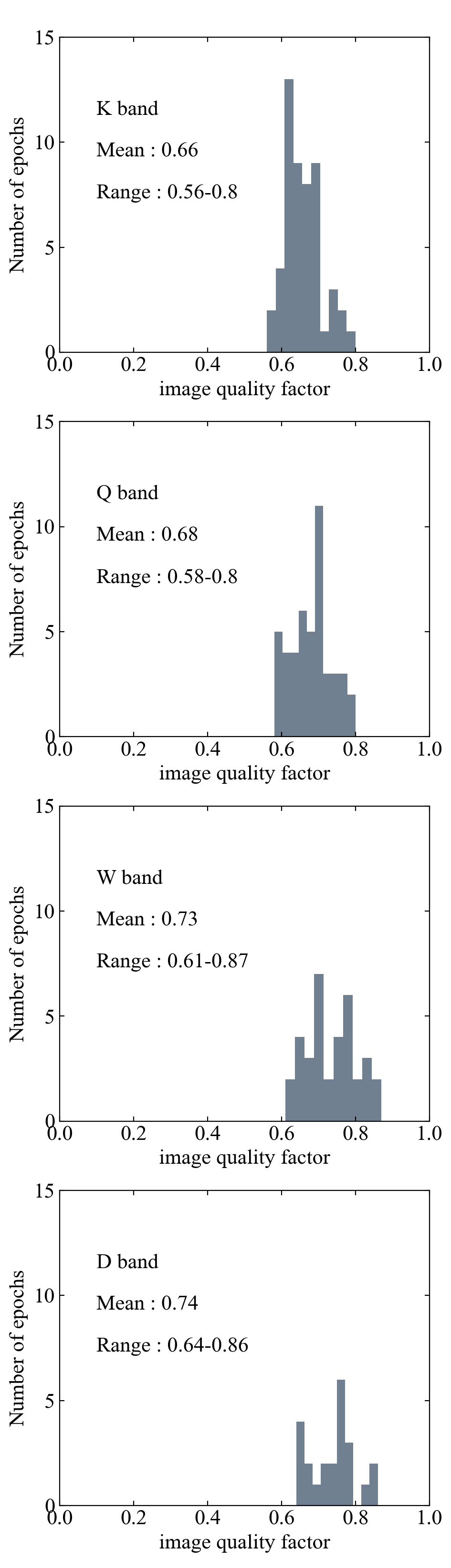}
\caption{Distributions of the image quality factor for the CLEANed images of 4C~+28.07 using the script.\label{fig:img_qual}}
\end{figure}

The ranges of the image quality factor are 0.56-0.76, 0.58-0.8, 0.61-0.87, and 0.64-0.86 at 22, 43, 86, and 129~GHz, respectively, and the corresponding mean values of the image quality factor are 0.65, 0.67, 0.74, 0.74 at 22, 43, 86, and 129~GHz. Detailed histograms are shown in Figure~\ref{fig:img_qual}. In addition, the image quality factor value for 3C~273 of KaVA shown in Figure~\ref{fig:clean_result_3c273} is 1.165 according to the script and 1.259 through manual CLEAN. We find that the CLEANed image from the script better describes the source brightness distribution.
%%% FIGURE %%%%%%%%%%%%%%%%%%%%%%%%%%%%%%%%%%%%%%%%%%%%%%%%%%%%%%%%%%
%%%%%%%%%%%%%%%%%%%%%%%%%%%%%%%%%%%%%%%%%%%%%%%%%%%%%%%%%%%%%%%%%%%%%

%%% CALL LIST OF REFERENCES (natbib STYLE) %%%%%%%%%%%%%%%%%%%%%%%%%%
% \bibliography{jkas-sample}

\end{document}